\documentclass[twocolumn]{aastex62}

\usepackage{lineno}
\usepackage{natbib}
\usepackage{subfigure}
\setcitestyle{sort,comma}
\usepackage{csvsimple}

\usepackage{amsmath}
\usepackage{fancyvrb}
\usepackage{multirow}
\usepackage{tabularx}

\newcommand\clearrow{\global\let\rowmac\relax}
\clearrow
\usepackage{graphicx}
\usepackage{array}
\usepackage{placeins}
\usepackage{hyperref}

\usepackage{siunitx}
\usepackage{color}
\usepackage{soul}

\usepackage{lineno}
\shorttitle{GJ 3929\,b - First Complete Rocky Worlds DDT Data Set}
\shortauthors{}

\begin{document}

\title{\large{GJ 3929\,b as the First Complete Rocky Worlds DDT Data Set}}

\correspondingauthor{Nicholas J. Connors}
\email{nicholas.connors@umontreal.ca}

\author[0000-0001-5848-6750]{Nicholas J. Connors} 
\affil{Department of Physics and Trottier Institute for Research on Exoplanets, Universit\'{e} de Montr\'{e}al, Montreal, QC, Canada}

\author[0009-0005-9152-9480]{Christopher Monaghan} 
\affil{Department of Earth, Planetary, and Space Sciences, University of California, Los Angeles, CA, USA}
\affil{Department of Physics and Trottier Institute for Research on Exoplanets, Universit\'{e} de Montr\'{e}al, Montreal, QC, Canada}

\author[0000-0001-5578-1498]{Bj\"{o}rn Benneke} 
\affil{Department of Earth, Planetary, and Space Sciences, University of California, Los Angeles, CA, USA}
\affil{Department of Physics and Trottier Institute for Research on Exoplanets, Universit\'{e} de Montr\'{e}al, Montreal, QC, Canada}

\author[0000-0003-4987-6591]{Lisa Dang}
\affil{Department of Physics and Trottier Institute for Research on Exoplanets, Universit\'{e} de Montr\'{e}al, Montreal, QC, Canada}
\affil{Department of Physics and Astronomy, University of Waterloo, Waterloo, Ontario, Canada, N2L 3G1}
\affil{Waterloo Centre for Astrophysics, University of Waterloo, 200 University W, Waterloo, Ontario, Canada, N2L 3G1}

\author[0000-0001-6809-3520]{Pierre-Alexis Roy} 
\affil{Department of Earth, Planetary, and Space Sciences, University of California, Los Angeles, CA, USA}
\affil{Department of Physics and Trottier Institute for Research on Exoplanets, Universit\'{e} de Montr\'{e}al, Montreal, QC, Canada}





\begin{abstract}

Despite their large abundance, it is still unknown whether and under what conditions rocky planets around M dwarf stars can host atmospheres. This open question motivated the ongoing Rocky Worlds Director's Discretionary Time survey focused on searching for atmospheres on relatively low temperature rocky exoplanets by systematically probing for the presence of day-night heat redistribution and CO$_2$ absorption through JWST/MIRI \SI{15}{\micro\meter} eclipse observations. Here we present the analysis of the first full data set from this survey, consisting of four observations of the warm Earth-size exoplanet GJ\,3929\,b, with a planetary mass of $1.75^{+0.44}_{-0.45}$~M$_\oplus$ and instellation flux of $17.3\pm0.7$~S$_\oplus$. In our analysis, we include two previously unpublished eclipse observations and find an overall eclipse depth of $118\pm22$~ppm and a dayside surface brightness temperature of $641^{+59}_{-64}$\,K. This is marginally lower than the eclipse depth of $160^{+26}_{-27}$ ppm previously reported based on only the first two observations. While the full data set remains consistent with bare rock scenarios, it also leaves more room for thin atmosphere scenarios. Only thick CO$_2$ atmospheres without thermal inversion remain ruled out at greater than 3$\sigma$. We also continue with lessons-learned in robustly analyzing these kind of high-precision JWST/MIRI \SI{15}{\micro\meter} eclipse observations. Notably, we find that the frame-normalized principal component analysis method appears more robust against the choice of extraction aperture size, which otherwise can have a significant impact on the inferred eclipse depth and scientific conclusions when using a standard polynomial baseline detrending method. 

\end{abstract}

\keywords{Exoplanets (498); Exoplanet atmospheres (487); Planetary atmospheres (1244)}


\section{Introduction} \label{sec:intro}

Rocky worlds orbiting M dwarfs are among the most common planets in our Galaxy, but their ability to retain atmospheres is still uncertain \citep{m_dwarf_frequency}. Habitable zone rocky worlds orbiting these stars are among the easiest to observe, due to the low luminosity and small radius of their host stars \citep{m_dwarf_hab_zone, ddt_working_group}. However, this proximity and the stellar activity of their host stars may hinder their ability to retain their atmospheres \citep{m_dwarf_atmo_loss}. By investigating the presence of atmospheres on these rocky planets at a population level, the ``cosmic shoreline" hypothesis, which describes a boundary between planets that can or cannot retain atmospheres based on their escape velocity and cumulative X-ray and ultraviolet (XUV) irradiation, can be empirically tested \citep{cosmicshoreline, ji2025cosmicshorelinerevisitedmetric, pass2025recedingcosmicshorelinemidtolate, bertathompson20253dcosmicshorelinenurturing}.

To distinguish between a planet with an atmosphere and a bare rock, we can measure the reduction in flux (eclipse depth) as the planet passes behind its host star. These close-in exoplanets are assumed to be tidally locked to their stars, meaning that only one side of the planet receives light from the star. This causes a large difference in temperature between the dayside and nightside if the planet has no atmosphere to redistribute heat between the two sides. The presence of an atmosphere could result in a lower measured dayside temperature and in a shallower eclipse depth due to advection of heat toward the nightside \citep{Koll_2022, coy2025populationlevelhypothesistestingrocky}. The eclipse depth can be further reduced by the presence of \SI{15}{\micro\meter} absorbing species such as CO$_2$ \citep{zieba_no_2023, greene_thermal_2023, august_hot_2024, hotrocks2, fortune2025hotrockssurveyiii, allen2025hotrockssurveyiv, 2026AJ....171..251H}. By performing secondary-eclipse observations, we are further able to avoid the effects of stellar contamination seen in transit observations \citep{rackham_2018_tls_effect}. The Mid-Infrared Instrument (MIRI) on board the James Webb Space Telescope (JWST) is uniquely suited to performing the mid-infra red observations of these targets \citep{MIRI}. By comparing the measured eclipse depth to surface and atmospheric models, we are able to infer their composition.

Previous studies of rocky exoplanets orbiting M dwarfs using \SI{15}{\micro\meter} MIRI photometry were focused on highly irradiated rocky worlds, and most were found to be bare rocks, with some tentative indications of atmospheres. TRAPPIST-1\,b was found to likely be a bare rock with a low surface albedo \citep{greene_thermal_2023}. A follow-up \SI{12.8}{\micro\meter} eclipse depth measurement may suggest a pure CO$_2$ atmosphere with photochemical hazes instead \citep{Ducrot_Lagage_Min_Gillon_Bell_Tremblin_Greene_Dyrek_Bouwman_Waters_etal._2025}, but atmospheric heat redistribution is not supported by the shape of its phase-curve \citep{trappist_phase_curve}. TRAPPIST-1\,c was found to be most consistent with a semireflective bare rock or a thin CO$_2$ and O$_2$ atmosphere \citep{zieba_no_2023}. LHS\,1140\,c was found to be consistent with a dark bare rock \citep{fortune2025hotrockssurveyiii, 2026ApJ...998L..39R}. TOI-1468\,b was found to be hotter than expected for a bare rock, implying either an additional source of heating, atmospheric thermal inversion, or uncharacterized detector systematic effects \citep{hotrocks2}. LHS\,1478\,b was given a tentative atmospheric detection based on only one visit, as the second visit was discarded due to no eclipse detection and significant systematic effects \citep{august_hot_2024} and will be followed up with additional JWST observations to confirm or reject this possibility in Cycle 4 \citep[GO-7675][]{2025jwst.prop.7675A}. Most recently, LTT\,3780\,b \citep{allen2025hotrockssurveyiv} and GJ\,3473\,b \citep{2026AJ....171..251H} were also found to be consistent with bare-rock compositions.

In order to further investigate the potential atmospheres of this population, the 500 hr Rocky Worlds Director's Discretionary Time (DDT) survey is ongoing, with a focus on more temperate planets to complement existing observations and further explore the parameter space spanned by the cosmic shoreline \citep{ddt_working_group}. The DDT survey will observe several exoplanets using MIRI \SI{15}{\micro\meter} secondary-eclipse photometry in order to detect or rule out the presence of a CO$_2$ atmosphere. The survey will also perform additional observations using the Hubble Space Telescope in order to better constrain the UV output of their stars.

The first target of the Rocky Worlds DDT is GJ\,3929\,b. It is an exo-Venus orbiting the M3.5V star GJ\,3929 ($T_{eff} = 3384\pm88$ K) alongside the nontransiting sub-Neptune mass planet GJ\,3929\,c, with a period of 2.6 days \citep{Beard_2022, Kemmer2022}. The planet has a radius of $1.09\pm0.04$~$R_\oplus$ and a mass of $1.75^{+0.44}_{-0.45}$~$M_{\oplus}$, with a surface equilibrium temperature of $568\pm6$~K and instellation flux of $17.3\pm0.7$~S$_\oplus$ \citep{Beard_2022}. The density of GJ\,3929\,b ($7.3\pm2.0$\,\SI{}{\gram / \cm^3}) does not suggest a thick atmosphere but leaves open the possibility that it could have a thin out-gassed atmosphere, a primordial volatile-poor atmosphere, or no atmosphere at all \citep{Beard_2022, seager_2010}. Of the nine initially announced DDT targets, GJ\,3929\,b has the lowest priority metric at $-0.2937$\footnote{\url{https://rockyworlds.stsci.edu/rw-website-targets.html}, retrieved June 4, 2026}, meaning it is among the least likely to have an atmosphere based on the cosmic shoreline hypothesis \citep{cosmicshoreline}.

In this paper we analyze four eclipses of GJ\,3929\,b using our open-source secondary-eclipse photometry pipeline \verb|Erebus| \citep{Connors2025}. This pipeline uses principal component analysis on the frame-normalized image time-series in order to isolate systematic effects. This serves as an independent follow-up to the analysis done on the first two eclipses in \mbox{\cite{xue2025jwstrockyworldsddt}}, and allows us to further investigate the systematic effects seen in previous MIRI \SI{15}{\micro\meter} observations.

This paper is divided into five sections. Section \ref{sec:data} details the data analysis methods used and the eclipse depth results. Section \ref{sec:systematics} discusses the systematic effects seen in the MIRI data, how they compare to previous \SI{15}{\micro\meter} secondary-eclipse photometry observations, and the impact of different aperture sizes and detrending methods. Section \ref{sec:scientific-results} discusses potential surface and atmospheric compositions of GJ\,3929\,b. Section \ref{sec:conclusion} summarizes our conclusions.

\section{Observations and Data Analysis}\label{sec:data}

\FloatBarrier

\begin{figure*}
    \centering
    \includegraphics[width=1\linewidth]{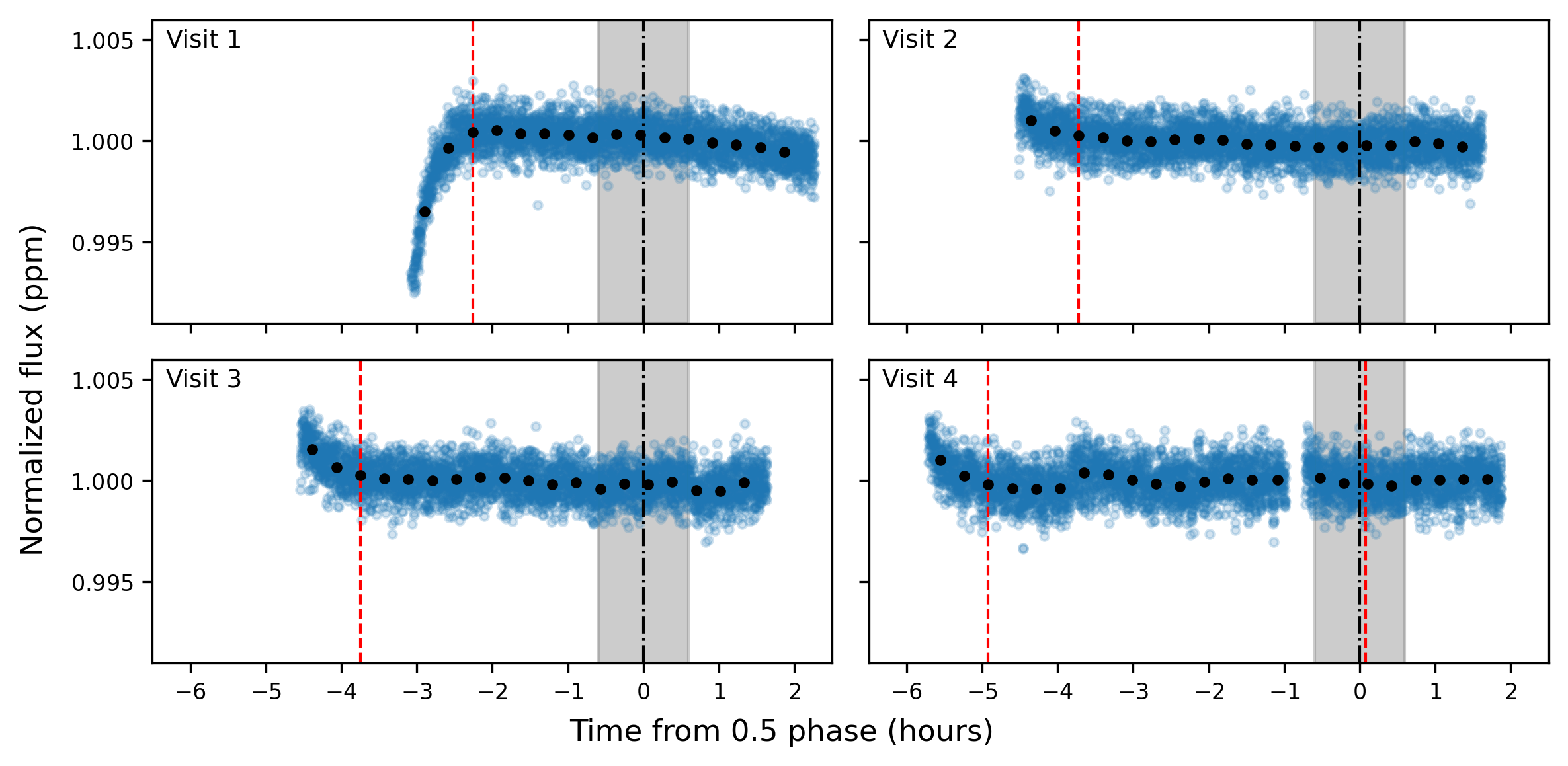}
    \caption{Raw light curves for the four visits of GJ\,3929\,b. The light curves shown use aperture photometry with an aperture radius of 5 pixels and a background annulus spanning 12-20 pixels. The extent of the detector-settling ramps for each observation are marked with red dashed lines (500 integrations). The grey shaded region shows the expected eclipse time for a circular orbit. Visit 4 consists of two observations, with a discontinuity shortly before the expected eclipse ingress. We see a possible tilt event in visit 4 around 3.5 hr before 0.5 phase.}
    \label{fig:raw_flux}
\end{figure*}

\begin{figure*}[t]
    \centering
    \includegraphics[width=\linewidth]{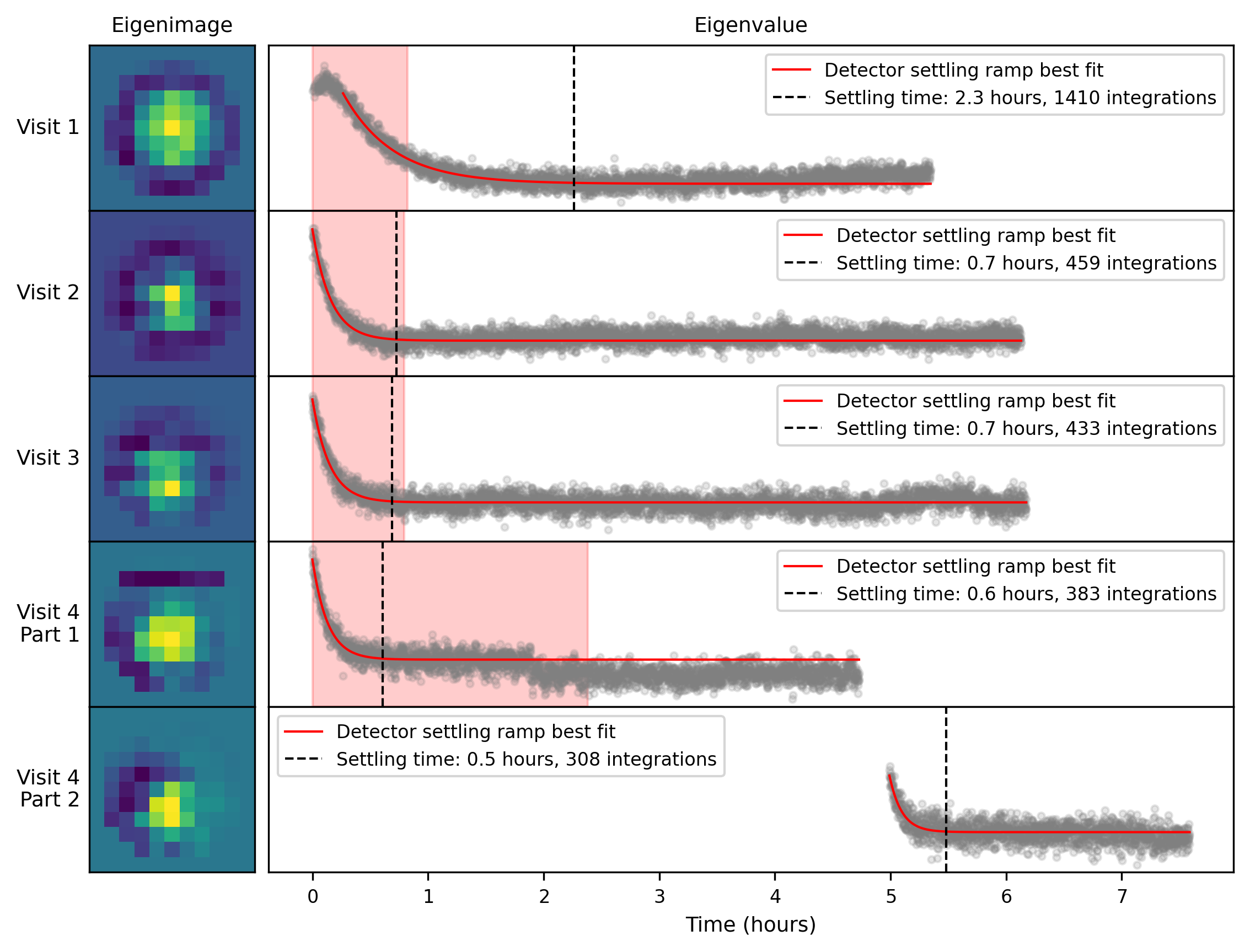}
    \caption{Detector-settling principal component eigenimages (left) and eigenvalues (right) for all visits of GJ\,3929\,b. The shaded region shows the range trimmed from the start of each observation. The first component features a lip that is not seen in the other observations and is not included in the exponential ramp fitting. The first visit component also shows an upward linear trend throughout the eigenvalue. The first observation of visit 4 has a sudden drop near the 2 hour mark, possibly marking a tilt event. The average settling times in hours are plotted against the $\text{K}_\text{s}$-band magnitude of the star in Figure \ref{fig:magnitude trend}.}
    \label{fig:settling_pca}
\end{figure*}

\begin{deluxetable}{c c c c c c}[t]
\tablecaption{Orbital parameters used to determine priors. \label{table:inputs}}
\startdata\\
$t_0$ (BJD-2,450,000) & $10452.8997\pm0.0012$ $^{(1)}$ \\
$P$ (days) & $2.6162644\pm0.0000026$ $^{(1)}$ \\
$R_p/R_*$ & $0.0318\pm0.0007$ $^{(1)}$ \\
$a/R_*$ & $17.05\pm0.43$ $^{(1)}$ \\
$i$ (deg) & $89.3\pm0.6$ $^{(1)}$ \\
$\Delta t_{sec}$ (hr) & $\mathcal{U}(-1.5, 0.5)$  \\
\enddata
\tablecomments{Reported uncertainties were used as standard deviations for Gaussian priors on relevant parameters. $\Delta t_{sec}$ is the time of secondary eclipse relative to 0.5 phase. $\mathcal{U}$ indicates a uniform prior. \\\textbf{References:} 
(1) \cite{xue2025jwstrockyworldsddt}}
\end{deluxetable}

GJ\,3929\,b was observed in secondary eclipse four times using MIRI: July 8th, 2025; July 31, 2025; January 30, 2026; and February 10, 2026. The observations were done using the F1500W filter with the SUB128 subarray and FASTR1 readout pattern. Each visit had 47 groups per integration, with 3387 integrations for visit 1 (6.1 hr), 3863 integrations (5.4 hr) for visit 2, and 3891 integrations (6.2 hr) for visit 3. Visit 4 was split into two non-interruptible observations required in order to observe for 7.1 hr. The first observation had 2980 integrations (4.7 hr) and the second observation had 1637 integrations (2.6 hr). The interruption between the two observations lasted 16 minutes and occurred close to 0.5 phase.

We begin by processing the uncalibrated MIRI data from the Mikulski Archive for Space Telescopes (MAST) labeled \verb|_uncal.fits| to stage 2 using the \verb|Eureka!| pipeline \citep{Bell2022}. In stage 1 we skip the EMI correction step (\verb!skip_emicorr = true! by default), and we skip the PHOTOM step (\verb!skip_photom = true!) in stage 2, such that the resulting data products can be used for absolute stellar flux calibration (see Section \ref{sec:stellarflux}). 

We then use \verb|Erebus| to perform aperture photometry of the calibrated data. We measure the flux within an aperture with varying sizes and use an annulus spanning 12 to 20 pixels to perform background subtraction. We mask out values marked \verb|DO NOT USE| and 3$\sigma$ outliers and replace them with 2d linearly interpolated data within the affected frame. 

We detrend the light curve using the frame-normalized principal component analysis (FN-PCA) technique described in \cite{Connors2025}. This technique consists of using principal component analysis on the time-series image frames of each observation, with each frame normalized such that we capture only interpixel variations in flux. We use a linear combination of the resulting principal component eigenvalues to remove systematic effects from the light curve. We also fit a linear slope to the data to model long-term stellar flux variations. We additionally repeat our analysis without the FN-PCA, instead using a third-degree polynomial fit as a baseline to compare against (previously used in light-curve detrending in e.g., \citealt{2026ApJ...998L..39R}, \citealt{zieba_no_2023}). We first attempted detrending using a linear slope and exponential ramp, but find that the third degree polynomial is required to be able to fully capture the shape of the light curve when not using FN-PCA. When using either method, we trim the start of the observation to mitigate the effect of the detector settling slope.

We use a \verb|batman| \citep{batman} model to represent the light curve and fit for it and our systematic model coefficients using a Markov Chain Monte Carlo (MCMC) with the \verb|emcee| library \citep{Foreman_Mackey_2013}. We test for convergence by running two MCMC chains in parallel and checking the Gelman-Rubin statistic \citep{gelman-rubin} and that the length of each chain is at least 50 times greater than the autocorrelation times of all parameters. We use the literature values presented in Table \ref{table:inputs} as Gaussian priors for the MCMC, along with a uniform prior from 0 to 2000 ppm for the eclipse depth. We test allowing negative values for the eclipse depth so as to not bias ourselves toward a detection, but due to the unconstrained eccentricity of the planet, this prevents us from finding the true time of the eclipse, instead fitting random out-of-eclipse baseline noise. Rather than doing this, we also fit for no eclipse and only our systematic model, and compare the goodness of this fit to determine whether a detection is preferred. We treat the orbit of the planet as if it were circular, but we fit for an offset to the secondary-eclipse time. We also perform a joint fit of all visits, where we bin the data from each visit by four in order to reduce computation time. In the joint fit the physical model parameters are shared between all visits, but each visit has its own set of parameters for the systematic model. Due to the strong detector settling ramps seen in the observations, we trim the first 500 integrations from the start of all observations, corresponding to the usual detector-settling length found in \cite{Connors2025}.

\subsection{Visit 1 Detector-settling Slope}

In visit 1 the FN-PCA is unable to fit the strong upward ramp seen at the beginning of the data (Figure \ref{fig:raw_flux}), requiring us to cut the data off before performing our fits. For this visit it is visually apparent that the detector-settling principal component does not match the shape of the ramp, due to the presence of an unexplained lip at the start of the eigenvalue time-series (Figure \ref{fig:settling_pca}). We speculate that this may be due to nonlinear systematic effects present in this visit that the FN-PCA is unable to fully capture.

\subsection{Visit 4 Observing Gap and Tilt Event}

The fourth eclipse was observed across two observations with a 16-minute gap separating them. To analyze this eclipse, we combine the two observations into a single light curve. The flux in each observation is separately normalized; otherwise, there is a large jump in the flux. Both observations are simultaneously detrended using the top 5 principal components of each. The linear slope in time is shared by both visits, with a step function between the two. There is a prominent ramp at the start of the second observation which is fit using an exponential, as the FN-PCA detrending is unable to fully capture it.

Visit 4 contains a prominent systematic error 2 hr after the start of the first observation (Figure \ref{fig:raw_flux}). At the same time, we see a sudden drop in the detector-settling principal component eigenvalue (Figure \ref{fig:settling_pca}) This bump in the light curve may be due to a tilt event. When fitting the light curve for this visit, we remove this event entirely by trimming the first 1500 integrations.

\subsection{Timing of Secondary Eclipse}\label{t_sec_timing}

The orbit of GJ 3929\,b is likely eccentric according to transit and radial velocity (RV) data, which indicate an eclipse timing offset of $-0.02\pm0.04$ days \citep{xue2025jwstrockyworldsddt}. From this, we use a uniform prior of $-0.5\pm1$ hr when performing the joint fits in this analysis. For the individual fits, we use the eclipse time found in the fiducial joint fit as a gaussian prior.

In our fiducial joint-fit analysis we find that the eclipse occurs earlier than expected for a circular orbit, further indicating that the orbit of GJ 3929\,b is slightly eccentric. The timing of the eclipse from our joint fit is $54\pm2$ minutes earlier than 0.5 phase. From this we are able to place a constraint on the eccentricity using

\begin{equation}
    e \cos \omega = \frac{\pi \Delta t}{2P}
\end{equation}

\noindent with orbital eccentricity $e$, argument of periastron $\omega$, period $P$, and eclipse offset from 0.5 phase $\Delta t$. From this we find $e\cos\omega = -0.0225\pm0.0010$.

We additionally perform a joint fit of the four visits with a shared eclipse depth but allow the timing of the eclipse to vary between visits. We find eclipse times of $-48\pm10$, $-57\pm4$, $-21\pm5$, and $-1\pm19$ minutes to 0.5 phase for the four visits. In visit 4, the fitted eclipse ingress corresponds to the detector-settling ramp caused by the observing gap. We compare the Bayesian information criterion (BIC; \citealt{schwarzBIC}) for the joint fits with and without a shared eclipse time and find that the shared eclipse timing offset is preferred. For our fiducial result we do not allow the eclipse time to vary between visits. 

\subsection{Eclipse Depth Result}

\begin{figure*}[t]
    \centering
    \includegraphics[width=1\linewidth]{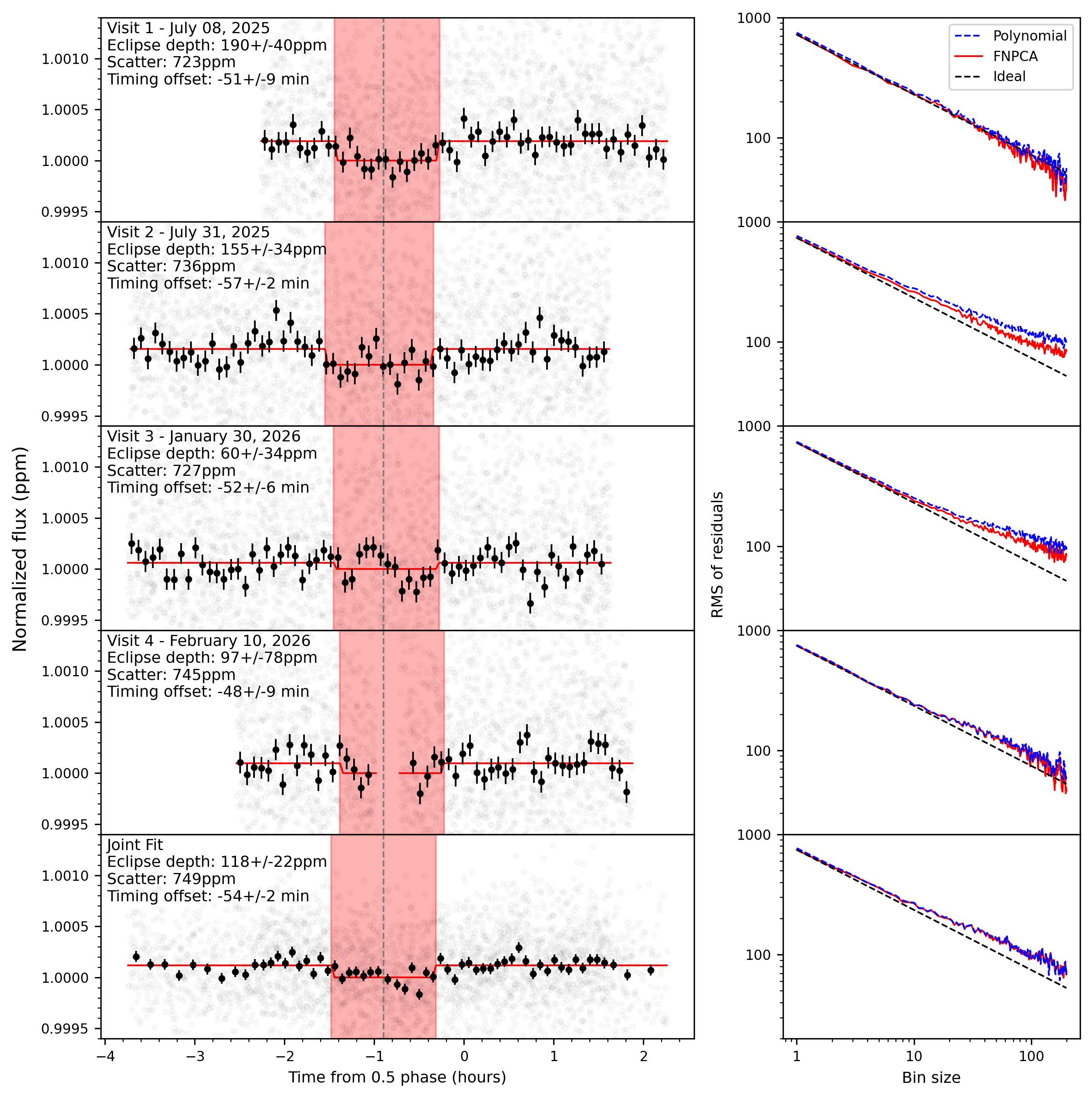}
    \caption{Individual and phase-folded joint-fit light curves of the four visits of GJ\,3929\,b found using FN-PCA detrending using the \texttt{Erebus} pipeline. The eclipse depth is found to be $190\pm40$\,ppm for visit 1, $155\pm34$\,ppm for visit 2, $60\pm34$\,ppm for visit 3, and $97\pm78$\,ppm for visit 4. There is a gap during the eclipse for visit 4 due to a scheduling constraint. For the joint-fit of the all visits the eclipse depth is $118\pm22$\,ppm. The dashed line shows the eclipse timing offset relative to a circular orbit retrieved from the joint fit ($54\pm2$\,minutes earlier than 0.5 phase). The eclipse time found in the joint fit was used as a gaussian prior on eclipse time for the individual fits. The RMS of residuals is plotted for both the FN-PCA and 3rd degree polynomial detrendings, showing it is generally improved by using the FN-PCA.}
    \label{fig:jointfit}
\end{figure*}

\begin{figure}[t]
    \centering
    \includegraphics[width=\linewidth]{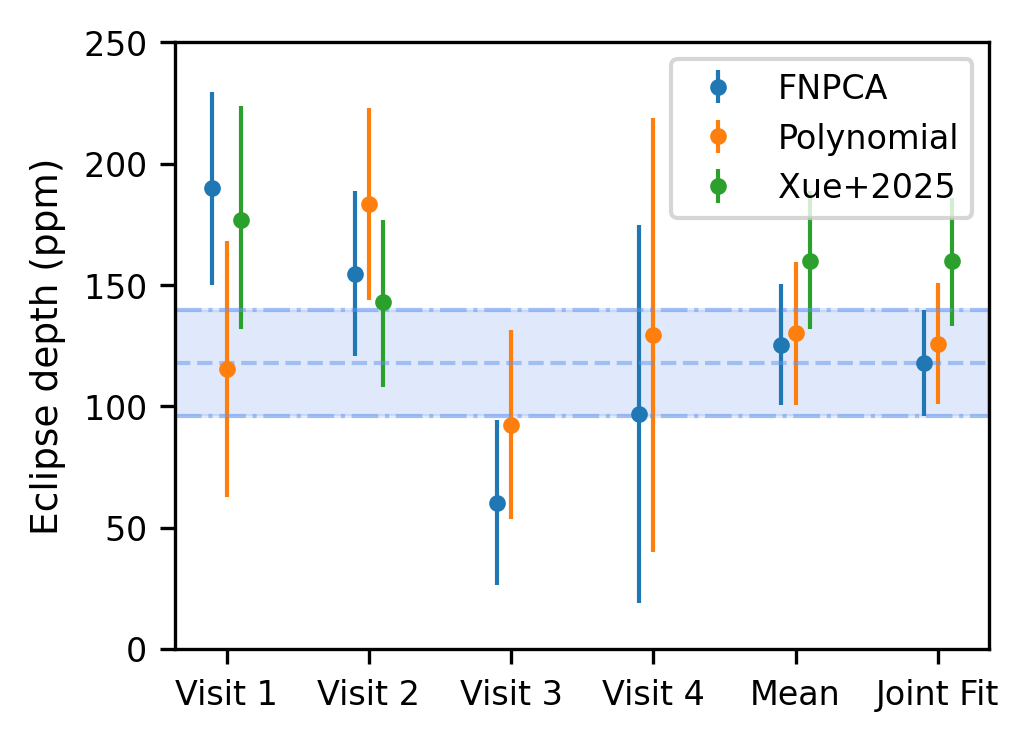}
    \caption{Individual and joint-fit results for the FN-PCA and linear detrending fits, along with the preliminary analysis of the first two released visits from \cite{xue2025jwstrockyworldsddt}. The shaded region marks the fiducial joint-fit result with error. The large error bars for Visit 4 are due to an observing gap occurring within the eclipse. The fitted FN-PCA light curves are shown in Figure \ref{fig:jointfit}.}
    \label{fig:eclipse depth}
\end{figure}

\begin{deluxetable}{ccc}[t]
\tablecaption{Eclipse depth results from different analyses. \label{tab:eclipse epth analysis}}
\startdata\\
 & \multicolumn{2}{c}{\textbf{Eclipse depth (ppm)}} \\
 \textbf{Analysis} & \textbf{FN-PCA} & \textbf{Polynomial} \\
\hline
Visit 1 & $190 \pm 40$ & $115\pm53$ \\
Visit 2 & $155 \pm 34$ & $183\pm40$ \\
Visit 3 & $60 \pm 34$ & $92\pm39$ \\
Visit 4 & $97 \pm 78$ & $129\pm89$ \\
Mean & $125 \pm 25$ & $130\pm28$ \\
Joint fit (all) & \colorbox{yellow}{$118 \pm 22$} & $126\pm25$\\
Joint fit (Excluding 1) & $96 \pm 26$ & - \\
Joint fit (Excluding 2) & $111 \pm 29$ & - \\
Joint fit (Excluding 3) & $146 \pm 26$ & - \\
Joint fit (Excluding 4) & $125 \pm 22$ & - \\
\enddata
\tablecomments{The FN-PCA detrending results use a 5 pixel aperture, the polynomial detrending results use a 6 pixel aperture. The "mean" analysis is the average of the individual visit fits. A leave-one-out analysis is done for each visit by performing a joint fit of the other visits using FN-PCA detrending. The fiducial joint-fit result is highlighted.}
\end{deluxetable}

We report a fiducial joint-fit eclipse depth of $118\pm 22$ ppm for the four visits of GJ 3929\,b using a 5 pixel aperture and FN-PCA detrending. This is in agreement with our best fit polynomial detrending result ($126\pm25$ ppm). Additionally, for aperture sizes $\geq5$, the joint-fit eclipse depths for the FN-PCA detrending are consistent within $1\sigma$. This is not the case for the polynomial detrending. We discuss the impact of aperture size choice in further detail in Section \ref{sec:aperture size}.

For our individual fits the mean eclipse depth is $125\pm25$ ppm when using the eclipse timing found in the fiducial joint fit. When allowing the timing of the eclipse to vary between visits but keeping a shared eclipse depth we find $137\pm23$ ppm. These results are in agreement with each other, although the latter is slightly deeper. If there are legitimate eclipse timing variations between visits, the fiducial joint fit would be including baseline in its measured eclipse depth for each visit, biasing the result towards a shallower value. The fitted lightcurves for the individual and shared eclipse time joint fit using FN-PCA are shown in Figure \ref{fig:jointfit}. We show a comparison of these results to those from the polynomial detrending and \cite{xue2025jwstrockyworldsddt} in Figure \ref{fig:eclipse depth}. We will use the result of the joint fit with a 5 pixel aperture, FN-PCA detrending, and shared eclipse time for the rest of the analysis in this paper.

We also perform a leave-one-out analysis to see the impact each individual visit has on the result of the joint fit. When excluding visit 1 from the joint fit (the deepest individual visit) we find an eclipse depth of $96\pm26$ ppm, about $1\sigma$ lower than our fiducial joint-fit result. When excluding visit 3 from the joint fit (the shallowest individual fit) we find an eclipse depth of $146 \pm 26$ ppm, about $1\sigma$ deeper than our fiducial joint fit. All values from this leave-one-out analysis agree within $<1.5\sigma$ of each other and the joint fit of all visits. The results of all individual and joint-fit analyses are shown in Table \ref{tab:eclipse epth analysis}. We note the joint-fit eclipse depth is virtually unchanged when excluding visit 4 given the poor quality of that visit.

We find variations in the eclipse depth up to $2.5\sigma$ between individual visits. This was previously found for LHS 1478\,b which had two secondary-eclipse observations with significant variation \citep{august_hot_2024}. A limitation of the FN-PCA approach is that it is unable to capture systematic effects that affect the overall throughput of the frame, such as those caused by stellar flaring. Variation between eclipses may then be due to unaccounted-for instrumental or astrophysical systematic effects like these. Eclipse depth variability has also been observed on 55 Cnc e, with suggested explanations including a 3:2 spin-orbit resonance \citep{2024A&A...690A.159P} and a transient outgassed atmosphere \citep{2023ApJ...956L..20H}. Our FN-PCA decomposition technique does not indicate anything that would explain these differences between eclipses.

\section{Instrumental Systematics and Detrending Techniques}\label{sec:systematics}

\begin{figure}[t]
    \centering
    \includegraphics[width=\linewidth]{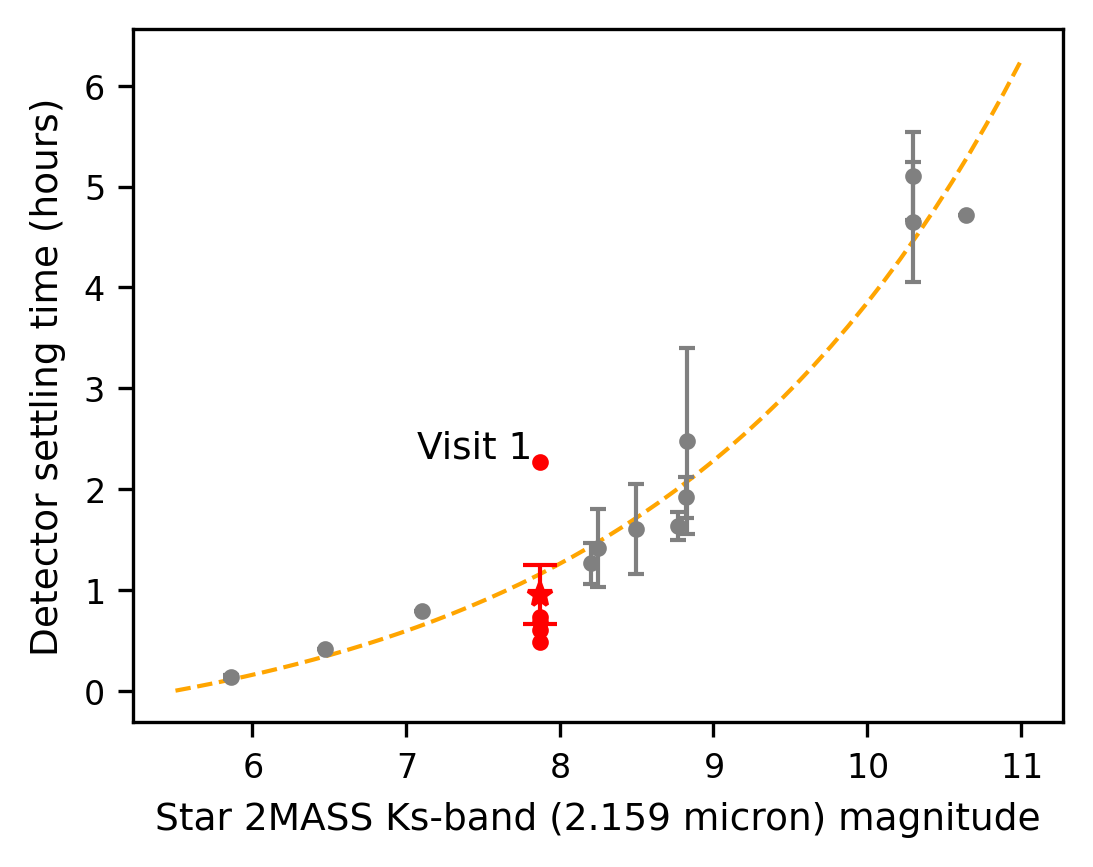}
    \caption{Detector-settling time from the five observations of GJ\,3929\,b in red (visit 4 is split into two). These values are found by fitting their detector-settling PCs with an exponential ramp and taking the time for 99\% of the signal to decay, giving 2.3, 0.7, 0.7, 0.6, and 0.5 hr for the five observations, respectively. The average settling time across all observations of GJ\,3929\,b is plotted as the red star, with the standard error. Visit 1 is labeled due to its notably high settling time. The dotted orange line is the detector-settling trend found in \cite{Connors2025}: $T_{settling} = 0.063e^{0.427m_{Ks}} -0.657$ hr.}
    \label{fig:magnitude trend}
\end{figure}

Here we investigate the instrumental systematics found in the data, the sensitivity to the chosen aperture size and detrending method, and the impact previous observations can have on detector settling. From our FN-PCA we analyze the top five principal components of the frame-normalized time-series image data of all five observations (visit 4 is split into two observations). We find that all five observations have highly ranked detector-settling principal components (Figure \ref{fig:settling_pca}). We find that the positive and negative diagonal centroiding components are present in the top 5 principal components for all observations as expected for a star of this brightness \citep{Connors2025}. The FN-PCA decomposition of all observations with and without trimmed data can be found in the Appendix. 

We measure the detector-settling time of each observation of GJ\,3929\,b by analyzing their detector-settling principal components. We fit the principal component eigenvalues with an exponential ramp and take the time for 99\% of the ramp to decay as the detector-settling time. We plot this settling time against the $\text{K}_\text{s}$-band magnitude of the star ($7.869\pm0.020$) and compare it to the detector settling trend previously discussed in \cite{Connors2025}, shown here in Figure \ref{fig:magnitude trend}. The time it takes these systematics to settle ($0.96\pm0.30$ hr) matches the previous trend. The average number of integrations required for it to settle, $599\pm183$, falls within error of the $525\pm203$ integrations required for detector settling found in \cite{Connors2025}.

\subsection{Sensitivity to Choice of Aperture Size}\label{sec:aperture size}

\begin{figure}
    \centering
    \includegraphics[width=\linewidth]{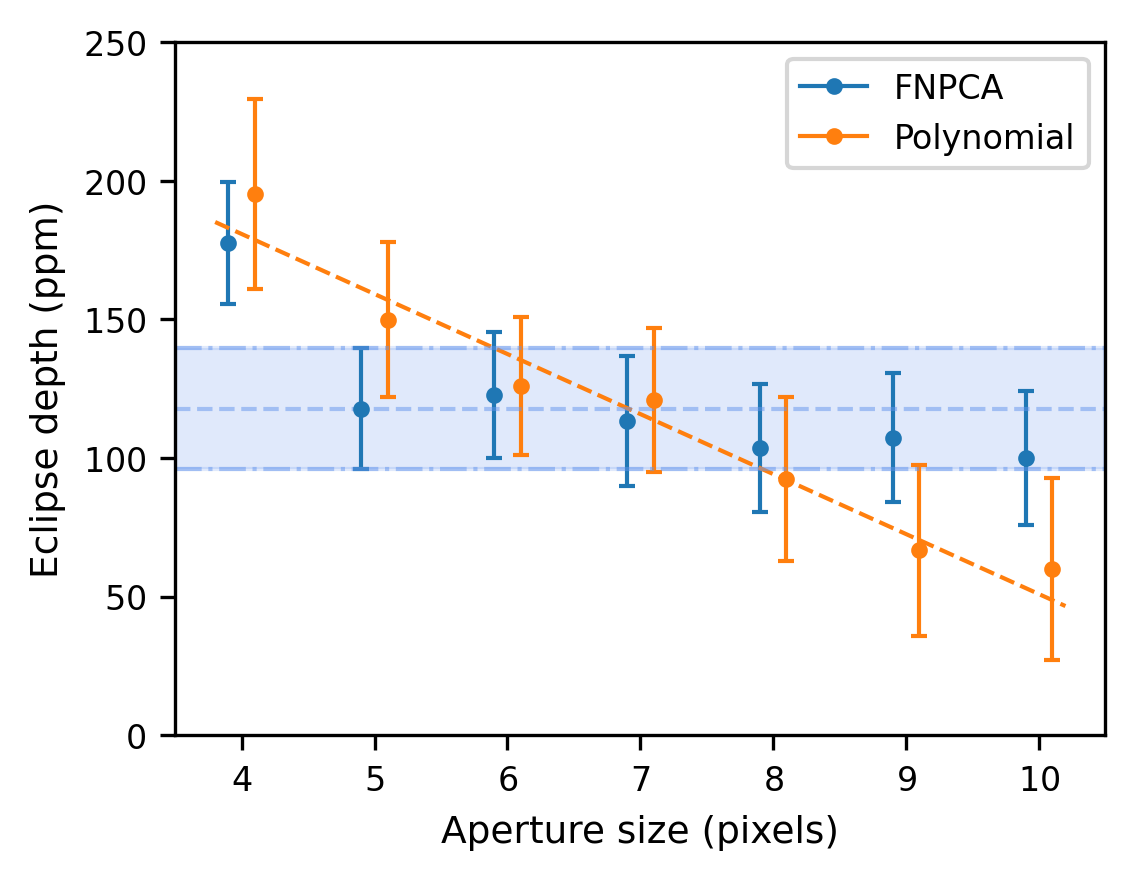}
    \caption{Joint-fit eclipse depth as a function of aperture radius for different detrending methods. The FN-PCA detrending with a 5-pixel aperture is chosen for the fiducial result, as it minimizes the BIC, and the shaded region marks this result with error. The data used to create this figure are shown in Table \ref{tab:aperture_size}. Our FN-PCA detrending is more stable under the choice of aperture size than the polynomial detrending, where the eclipse depth decreases linearly with aperture size.}
    \label{fig:FN-PCA_aperture_size}
\end{figure}

\begin{deluxetable}{cccc}[t]
\tablecaption{Joint fit results for different aperture sizes. \label{tab:aperture_size}}
\startdata\\
    \multicolumn{4}{c}{FN-PCA detrending} \\
    Radius (px) & $f_p/f_\star$ (ppm) & $\Delta t_{sec}$ (minutes) & $\Delta$ BIC \\
    \hline
        4 & $177 \pm 22$ & $-56 \pm 0.6$ & 329 \\
        \colorbox{yellow}{5} & \colorbox{yellow}{$118 \pm 22$} & \colorbox{yellow}{$-54 \pm 2.4$} & \colorbox{yellow}{0} \\
        6 & $123 \pm 23$ & $-53 \pm 2.3$ & 102 \\
        7 & $113 \pm 23$ & $-53 \pm 2.3$ & 191 \\
        8 & $104 \pm 23$ & $-56 \pm 2.7$ & 213 \\
        9 & $107 \pm 23$ & $-56 \pm 2.5$ & 206 \\
        10 & $100 \pm 24$ & $-55 \pm 3.4$ & 294 \\\hline
    \multicolumn{4}{c}{Polynomial detrending} \\
    Radius (px) & $f_p/f_\star$ (ppm) & $\Delta t_{sec}$ (minutes) & $\Delta$ BIC \\\hline
        4 & $195 \pm 34$ & $-57 \pm 1.3$ & 2402 \\
        5 & $150 \pm 28$ & $-56 \pm 1.1$ & 673 \\
        \colorbox{yellow}{6} & \colorbox{yellow}{$126 \pm 25$} & \colorbox{yellow}{$-53 \pm 2.5$} & \colorbox{yellow}{56} \\
        7 & $121 \pm 26$ & $-53 \pm 2.4$ & 165 \\
        8 & $92 \pm 30$ & $-54 \pm 9.4$ & 302 \\
        9 & $67 \pm 31$ & $-51 \pm 17.8$ & 486 \\
        10 & $60 \pm 33$ & $-25 \pm 11.5$ & 771 \\\hline
\enddata
\tablecomments{$\Delta t_{sec}$ is defined as the time of eclipse compared to 0.5 phase. We use the BIC to compare the goodness of fit of both detrending methods (FN-PCA or third-degree polynomial). The precision of the eclipse timing is lessened with increasing aperture size, but at a notably faster rate for the polynomial detrending than the FN-PCA. The best fit results are highlighted.} 
\end{deluxetable}

When performing aperture photometry, a variety of criteria can be employed to select an apparently optimal aperture size. Here we try a range of different aperture sizes from 4 to 10 pixels to study how this choice affects the resulting eclipse depth. We perform joint fits of all four visits using our two detrending methods with these different apertures.

We find that the measured eclipse depth varies as a function of aperture size (Figure \ref{fig:FN-PCA_aperture_size}). For the FN-PCA detrending this effect is most notable when using the 4-pixel aperture. This is likely due to the detector-settling systematic affecting more of the smaller aperture, as this systematic is most prominent in the center of the point-spread function. For an aperture size of 5 pixels and above, the resultant eclipse depth is more stable as a function of aperture size, with all eclipse depths agreeing within $1\sigma$. However, we do observe a slight downward trend. For our baseline polynomial detrending, the resultant eclipse depth is linearly dependent on our choice of aperture size. Additionally, we find that the uncertainty in the timing of the eclipse increases with the chosen aperture size when using the polynomial detrending, and less prominently when using FN-PCA (Table \ref{tab:aperture_size}).

When comparing the FN-PCA and polynomial detrending methods, we find that the former is more robust against the choice of aperture size. For our fiducial eclipse depth result, we use the 5-pixel FN-PCA result, which is the smallest aperture size where the eclipse depth is stable. The smaller aperture size also means that it includes less background noise, and it also minimizes the BIC. The dependence of the eclipse depth on the aperture size stems from different apertures changing the overall shape of the out-of-eclipse baseline, as additional pixels are incorporated into the light curve. The changing shape of the baseline can result in entirely different eclipse depths being measured. For the polynomial case, this difference can result in an entirely different inferred composition for the planet, from a dark bare rock using a 5-pixel aperture to a thick atmosphere using 10 pixels. The FN-PCA technique is performed on all pixels within the chosen aperture, allowing it to innately capture the variation within these added pixels. From this, we are more able to robustly measure the eclipse depth. 

We note that choosing a smaller aperture size would seemingly give a joint-fit result in agreement with the conclusion of \cite{xue2025jwstrockyworldsddt} ($160^{+26}_{
-27}$ ppm). However, when considering only the individual fit eclipse depths of the first two visits that were analyzed in this previous work, we find good agreement between their fiducial result and our 5-pixel FN-PCA result (Figure \ref{fig:eclipse depth}). Additionally, the shallower result is largely stable under the choice of aperture size, unlike the 4-pixel aperture result.

\subsection{Effect of Previous Filter}

Before slewing to the target for the four visits of GJ\,3929\,b, the MIRI instrument was changed to its prism mode (filter P750L). This was done in order to test the hypothesis presented in \cite{fortune2025hotrockssurveyiii} that the previous filter may impact the slope of the detector-settling ramp\footnote{\url{https://rockyworlds.stsci.edu/downloads/GJ3929b_JWST_SchedulingReport_2025-08-18.pdf}, retrieved 2025 September 8}. In Figure 12 of \cite{fortune2025hotrockssurveyiii}, observations where P750L was the previous filter have either flat or slightly negative settling slopes. 

\begin{figure}[t]
    \centering
    \includegraphics[width=0.95\linewidth]{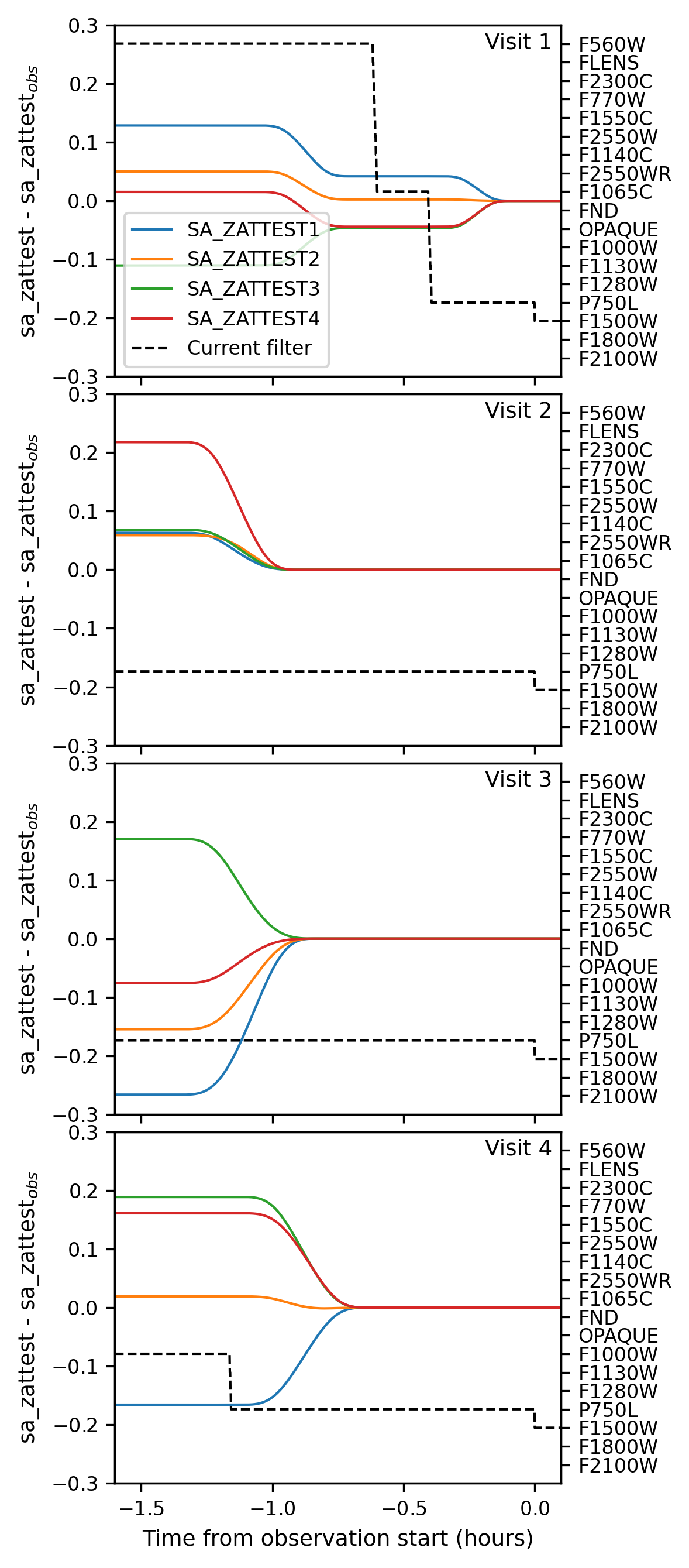}
    \caption{Engineering mnemonics for the four visits of GJ\,3929\,b prior to observation. The \texttt{sa\_zattest} engineering mnemonics indicate the pointing of the telescope, with the y-axis arranged such that the telescope is on target when the curves converge to zero. The filters are ordered to accurately reflect the rotation of the filter wheel. The pre-flashing detector-settling slope experiment has the filter change to P750L before slewing to the target. The first visit is discrepant in that it spends very little time in this filter, which may impact the systematics present in the rest of the visit.}
    \label{fig:apendix engineering mnemonics}
\end{figure}

We compare the slope of the first 30 minutes of the four visits of GJ\,3929\,b to those shown in Figure 12 of \cite{fortune2025hotrockssurveyiii}. We follow the procedure outlined in their paper by taking the light-curve slopes of the four visits using a linear best fit through the first 30 minutes of data and reporting it as a percentage per hour. We find a slope of $1.35\pm0.04$ \%/hour for visit 1, $-0.22\pm0.03$ for visit 2, $-0.28 \pm 0.03$ for visit 3, and $-0.38 \pm 0.03$ for visit 4. We do not consider the size of the detector-settling slope for the fifth observation as it was performed without the pre-flashing experiment. We find that visits 2-4 have slopes whose sizes are clustered together, while visit 1 is significantly discrepant. Excluding the first visit, these slopes are similar to those seen in \cite{fortune2025hotrockssurveyiii} when the P750L filter had been previously used. When investigating the engineering mnemonics using \verb|spelunker| \citep{Deal2024} we find that visit 1 was on target in the P750L filter for under 10 minutes, while the other visits of GJ\,3929\,b were on target for 30 to 45 minutes before starting the observation (Figure \ref{fig:apendix engineering mnemonics}). A similar discrepancy between observations that used the same filter previously was seen in \cite{fortune2025hotrockssurveyiii}, where observations of GJ\,3473\,b and LHS\,1478\,b had significantly different settling times after being on target in the F560W filter for 45 minutes and 10 minutes, respectively. Additionally, the discrepant visit of GJ\,3929\,b was only in the P750L filter for 24 minutes total prior to the beginning of the observation, including the time spent slewing to the target. The other observations that had previously used the P750L filter (both those originally reported in \cite{fortune2025hotrockssurveyiii} and those reported in this work) all spent at least a full hour in this filter and have similar settling slopes. We have included figures showing the relationship between settling slope and time on P750L in the Appendix.

We note some important caveats in this comparison: 1) The trend in the Hot Rocks Survey is for the 256 sub-array, but this data was taken with the 128 sub-array; 2) The original trend was observed when using different pipeline and data reduction choices. In order to more accurately compare our results to those of the Hot Rocks Survey, we have recalculated the detector settling slopes of the targets which had been in the P750L filter prior to observation with the same aperture photometry used for GJ\,3929\,b in this paper. We see that the magnitude of the detector settling slope may be impacted by both the previous filter and the time spent in that filter both on and off target.

\section{Atmosphere and surface analysis}\label{sec:scientific-results}

\begin{figure*}[t!]
    \centering
    \includegraphics[width=1\linewidth]{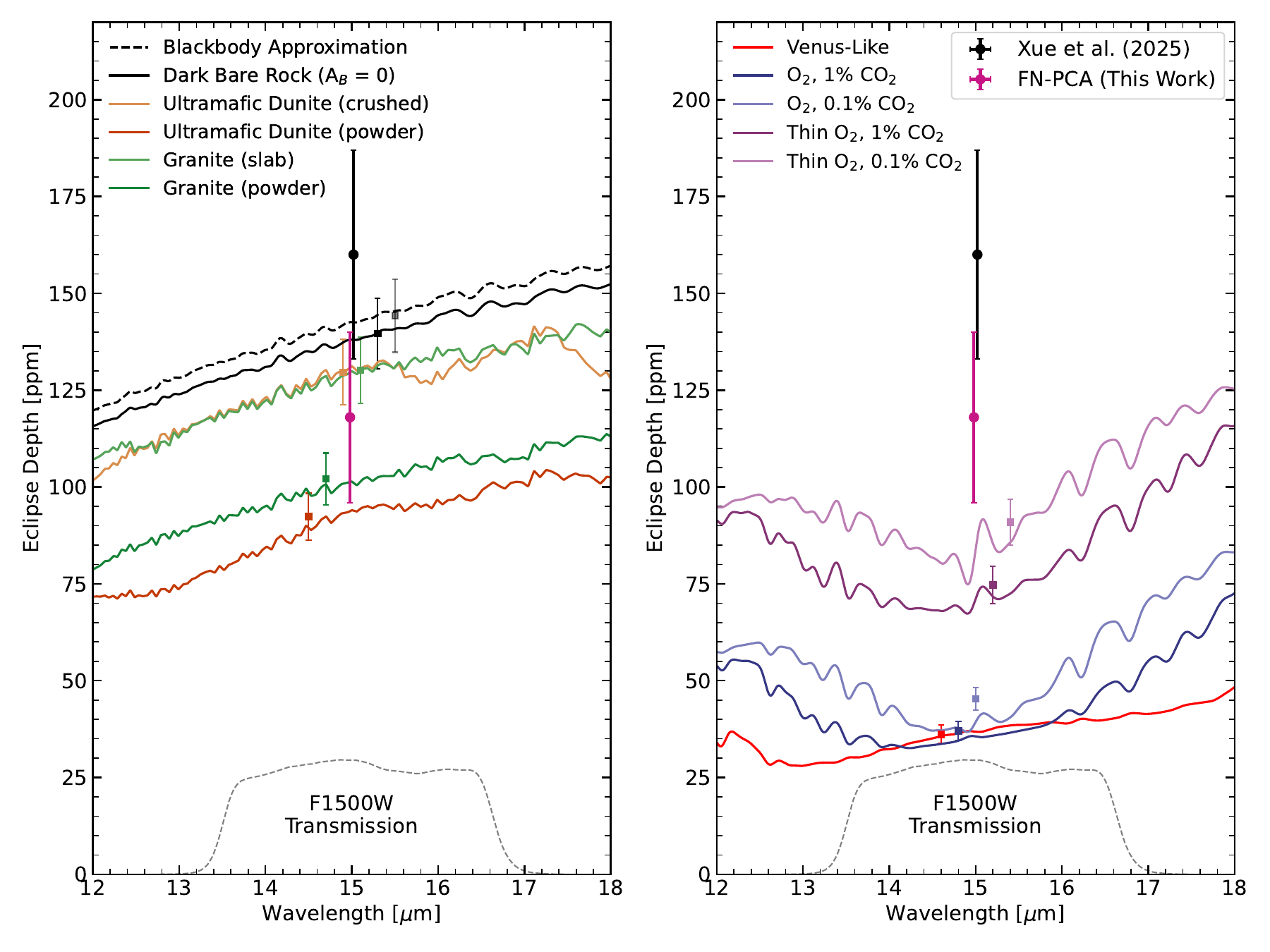}
    \caption{Simulated emission spectra of GJ\,3929\,b for various surface (left) and atmospheric (right) compositions compared to the measured eclipse depths from our FN-PCA calculation, alongside the previously reported result using only the first two visits by \citet{xue2025jwstrockyworldsddt}. The F1500W filter transmission is shown with the dotted line at the bottom. The simulated MIRI 1500W eclipse depths for each of the models are shown as squares in each panel, with error bars indicating the estimated model uncertainty (Section \ref{sec:modeluncertainty}; \citealt{2026arXiv260415421M}). Small horizontal offsets between points are added for clarity. The ultramafic dunite, granite, and basalt model spectra were generated using the dunite xenolith, dalmatian granite, and K1919 basalt spectral data from \citet{Paragas_Knutson_Hu_Ehlmann_Alemanno_Helbert_Maturilli_Zhang_Iyer_Rossman_2025}. The thin-atmosphere models were generated assuming zero heat redistribution to the nightside, and the thick-atmosphere models assume full heat redistribution. Additional models and $\sigma$ confidences are reported in Table \ref{tab:models}.}
    \label{fig:surface_models}
\end{figure*}

We simulate emission spectra of GJ\,3929\,b using different bare-rock and atmospheric models (Figure \ref{fig:surface_models}) as was done in our uniform reanalysis of MIRI \SI{15}{\micro\meter} observations in \cite{Connors2025} to infer the presence of an atmosphere. We simulate the flux of the host star ($f_\star$) using a SPHINX stellar model \mbox{\citep{Iyer_Line_Muirhead_Fortney_Gharib-Nezhad_2023, Iyer_Line_Muirhead_Fortney_Gharib-Nezhad_2024}} in order to convert from the measured eclipse depth ($f_p/f_\star$) to the planetary flux ($f_p$). To ensure that the $f_\star$ value is accurate at \SI{15}{\micro\meter} we perform an absolute stellar flux calibration based on the measured flux from the star during the eclipse \citep{Gordon_2025}. 

\subsection{Atmospheric Modeling}
We use the \verb|SCARLET| modeling framework to create 1D self-consistent atmospheric models to model the CO$_2$ absorption feature at \SI{15}{\micro\meter} \citep{scarlet_1, scarlet_2, scarlet_3, 2019ApJ...887L..14B, 2019NatAs...3..813B, benneke2024jwstrevealsch4co2, 2021AJ....162...73P, Pelletier_Benneke_Chachan_Bazinet_Allart_Hoeijmakers_Lavail_Prinoth_Coulombe_Lothringer_et_al._2024, Roy_Benneke_Piaulet_Crossfield_Kreidberg_Dragomir_Deming_Werner_Parmentier_Christiansen_et_al._2022, Roy_2023, Piaulet_Ghorayeb_2024, Bazinet_2024, Monaghan_Roy_Benneke_Crossfield_Coulombe_PiauletGhorayeb_Kreidberg_Dressing_Kane_Dragomir_etal._2025}. These models assume a nongrey, radiative, convective temperature profile and a well-mixed composition of CO$_{2}$ in an O$_{2}$ dominated atmosphere and solve for hydrostatic equilibrium and radiative transfer iteratively to produce a theoretical temperature-pressure profile. Once the model converges to stability, the associated emission spectrum as measured from the secondary eclipse is computed. The Venus-Like model assumes a composition of 96.5\% CO$_{2}$ and 3.5\% O$_{2}$. The surface pressure of each model is set to be 30 mbar for the thin atmosphere models and 100 mbar otherwise. The thin models are simulated without heat recirculation ($f = 2/3$), implying that incident flux from the host star is not redistributed uniformly across the planet's surface, resulting in a hot dayside and cool nightside. The thick atmospheres are simulated with full heat recirculation ($f = 1/4$), such that the temperature across the planet's surface is roughly uniform owing to the presence of a thick, redistributive atmosphere.

\subsection{Surface modeling}
We simulate a range of bare-rock emission spectral models representing different possible surface albedos using \verb|JESTER|. This tool separates the dayside hemisphere into a number of annular sections to properly account for the localized energy balance \citep{Monaghan_Roy_Benneke_Crossfield_Coulombe_PiauletGhorayeb_Kreidberg_Dressing_Kane_Dragomir_etal._2025, 2026arXiv260415421M}. The wavelength-dependent hemispherical reflectance of lab samples measured in \citet{Paragas_Knutson_Hu_Ehlmann_Alemanno_Helbert_Maturilli_Zhang_Iyer_Rossman_2025} was used to calculate the emission spectra of GJ\,3929\,b as an ultramafic and granite surface. Two different grain sizes are simulated for each model. In addition to these models, a dark bare rock with $A_{\mathrm{B}}$ = 0 and a uniform temperature blackbody are also generated for comparison. 

\subsection{Absolute Stellar Flux Calibration}
\label{sec:stellarflux}

To generate the atmospheric and surface models described above, we model the host star's flux $f_\star$ using an interpolated SPHINX stellar model \citep{Iyer_Line_Muirhead_Fortney_Gharib-Nezhad_2023, Iyer_Line_Muirhead_Fortney_Gharib-Nezhad_2024}. We assume the host star's parameters follow those measured in \cite{Kemmer2022} for the effective temperature, log($g$), and stellar metallicity. Furthermore, we assume a C/O value of 0.56 from the mean distribution of existing C/O ratios previously reported in other M dwarf stars \citep{2014PASJ...66...98T, 2016PASJ...68...13T, 2015PASJ...67...26T, Nakajima_Sorahana_2016}.

Our value for the flux of the planet, $f_p$, is based on the measured eclipse depth ($f_p/f_\star$) and a model stellar flux ($f_\star$) value. To ensure this $f_\star$ value is accurate at \SI{15}{\micro\meter}, we follow the procedure outlined in \mbox{\cite{Gordon_2025}} and \cite{fortune2025hotrockssurveyiii}. As specified in \cite{Gordon_2025} for the F1500W filter we use an aperture size of 5.69, and a background annulus spanning 8.63 to 11.45 pixels using \verb|photutils| \mbox{\citep{larry_bradley_2024_13989456}}. We convert from DN/s/pixel to surface brightness MJy/sr using Equation 3 in \cite{Gordon_2025} with the aperture correction to an infinite aperture $A_{cor} = 1.497\pm0.0190$ for F1500W, and the average solid angle per pixel $\Omega_{pix} = 0.11$\,arcsec/pixel. 

We find that the measured stellar flux during the eclipse across all visits is 19.19 $\pm$ 0.12 mJy. The MIRI 1500W stellar flux generated by our SPHINX models measures to be 18.8 mJy, and we thus scale our generated eclipse depths such that the SPHINX model matches the measured stellar flux.

\subsection{Model Uncertainty}
\label{sec:modeluncertainty}

The imprecision of different astrophysical parameters such as the ratio of planet to star radii $\frac{R_{\mathrm{p}}}{R_{\mathrm{*}}}$ and the planet's scaled semi-major axis $a_{\mathrm{p}}/R_{\mathrm{*}}$ will inject some level of uncertainty into the accuracy of our forward models. Furthermore, even with our stellar flux calibration, the accuracy of the model spectrum for the host star remains difficult to determine without further observations of GJ 3929 \citep{Fauchez_2025}. Following \citet{2026arXiv260415421M}, we estimate the model uncertainty of GJ 3929\,b in the MIRI F1500W filter to be $6.6\%$ by accounting for uncertainty in $\frac{R_{\mathrm{p}}}{R_{\mathrm{*}}}$, $a_{\mathrm{p}}/R_{\mathrm{*}}$, and stellar properties that modify the outgoing flux from the host star. 

\subsection{Results}

\begin{deluxetable}{ccc}[t]
\tablecaption{Comparison to surface and atmosphere models \label{tab:models}}
\startdata\\
 \textbf{Model} & \textbf{Eclipse depth} & \textbf{$\sigma$ confidence} \\
 & \textbf{(ppm)} & \\
\hline
    Venus-Like, 100 mbar & $36.1\pm 2.4$ & 3.7 \\
    O$_{2}$, 1\% CO$_{2}$, 100 mbar & $37.1\pm2.4$ & 3.7 \\
    O$_{2}$, 0.1\% CO$_{2}$, 100 mbar & $45.4\pm 3.0$ & 3.3 \\
    Thin O$_{2}$, 1\% CO$_{2}$, 30 mbar & $74.7\pm 4.9$ & 1.9 \\
    Thin O$_{2}$, 0.1\% CO$_{2}$, 30 mbar & $91.0 \pm 6.0$ & 1.2 \\
    Ultramafic Dunite (powder) & $92.3\pm 6.0$ & 1.1 \\
    Granite (powder) & $102.1 \pm 6.7$ & 0.69 \\
    Ultramafic Dunite (crushed) & $129.6 \pm 8.5$ & 0.49 \\
    Granite (slab) & $130.1 \pm 8.5$ & 0.51 \\
    Dark Bare Rock (A$_{B}$ = 0) & $139.7 \pm 9.1$ & 0.91 \\
    Blackbody Approximation & $144.2 \pm 9.4$ & 1.10 \\
\enddata
\tablecomments{Eclipse depths are reported by integrating over the MIRI F1500W bandpass. The "$\sigma$ confidence" column is the certainty with which we rule out this composition compared to the fiducial eclipse depth $118\pm22$ ppm. A selection of models are plotted in Figure \ref{fig:surface_models}.}
\end{deluxetable}

\begin{figure}
    \centering
    \includegraphics[width=\linewidth]{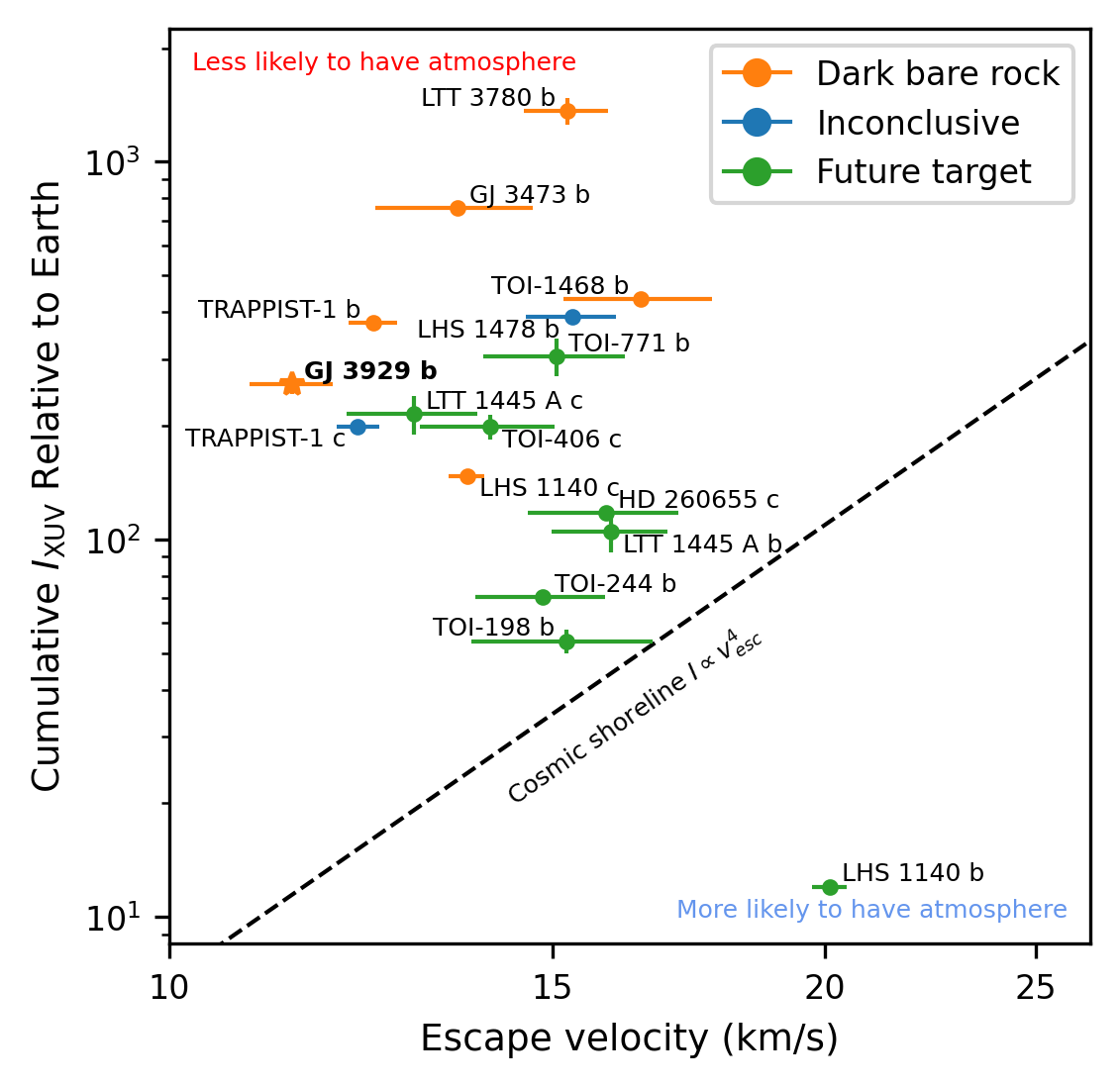}
    \caption{Cosmic shoreline plot of exoplanets previously observed at 15 $\mu$m using MIRI alongside the remaining DDT targets \citep{greene_thermal_2023, zieba_no_2023, august_hot_2024, hotrocks2, fortune2025hotrockssurveyiii, allen2025hotrockssurveyiv, 2026AJ....171..251H}. The cosmic shoreline is shown as the line $I \propto v_{\mathrm{esc}}^4$ such that it intersects the $I_{\mathrm{XUV}}$ and $v_{\mathrm{esc}}$ for Mars using equation 27 in \cite{cosmicshoreline}. Data used to generate this figure is from \cite{2024ApJ...960L...3C, 2019AJ....158..138S, 2023A&A...675A..52C,
2023AJ....165..134O, 2023AJ....166..171P, 2022AJ....163..168W,
Beard_2022, 2024arXiv240911083L, 2021PSJ.....2....1A,
2024A&A...682A..66B, 2021A&A...649A.144S, 2025A&A...698A..68M,
2022A&A...664A.199L, 2026A&A...706A.166Z, 2020A&A...642A.236K}.}
    \label{fig:cosmicshoreline}
\end{figure}

We find that the measured eclipse depth of GJ 3929\,b is most consistent with a bare rock as predicted by the cosmic shoreline model (Figure \ref{fig:cosmicshoreline}). Given the observational and model uncertainty, the measured eclipse depth is most consistent with the atmosphere-depleted, low surface albedo models simulated by \texttt{JESTER}. However, we cannot statistically rule out any surface or thin-atmospheric composition at greater than $3\sigma$ with the current observations. To further characterize the surface composition would require observations at other wavelength ranges. We are only able to rule out thick atmospheres with complete heat recirculation and the presence of CO$_2$. Our simulated models and the $\sigma$ confidence with which they are ruled out by our observations are compiled in Table \ref{tab:models}.

We calculate the dayside brightness temperature $T_{\mathrm{d}}$ of GJ\,3929\,b from the FN-PCA measured eclipse depth using a simple minimization function. The photon flux $f_{\gamma}$ measured in the F1500W filter from the planet (assuming it emits as a uniform temperature blackbody) and host star is calculated as:

\begin{equation}
    f_{\gamma,\,\mathrm{p}} = \int \frac{\pi B_{\mathrm{p}}(T_{\mathrm{d}}) \lambda}{hc} W_{\lambda}\,d\lambda
\end{equation} \label{eq:photonfluxplanet}
\begin{equation}
    f_{\gamma,\,\mathrm{*}} = \int \frac{f_\star \lambda}{hc}W_{\lambda}\, d\lambda
\end{equation}\label{eq:photonstarplanet}

\noindent where $B_{\mathrm{p}}(T_{\mathrm{d}})$ represents the Planck function and $W_{\lambda}$ the throughput of the F1500W filter. For each planet, we find the value $T_{\mathrm{d}}$ such that $\frac{R_{\mathrm{p}}^{2}}{R_{\mathrm{*}}^{2}}\frac{f_{\gamma,\,\mathrm{p}}(T_{\mathrm{d}})}{f_{\gamma,\,\mathrm{*}}}$ is equal to the FN-PCA measured eclipse depth. From these equations, we calculate a dayside brightness temperature of 641$^{+59}_{-64}$\,K, which is $\sim2\sigma$ less than the temperature previously reported by \citet{xue2025jwstrockyworldsddt}.

We may contextualize our measured brightness temperature by calculating the brightness temperature ratio:

\begin{equation}
    \mathcal{R} = T_{\mathrm{d}}/T_{\mathrm{d,max}}
\end{equation}

where $T_{\mathrm{d,max}}$ represents the expected dayside temperature for a dark bare rock with $A_{\mathrm{B}} = 0$. For a planet consistent with a dark bare rock, we anticipate that $\mathcal{R} = 1$. Assuming that $ T_{\mathrm{max}} = T_{\mathrm{*}} \sqrt{\frac{R_{\mathrm{*}}}{a_{\mathrm{p}}}}\left(\frac{2}{3}\right)^{\frac{1}{4}}$ \citep{2008ApJS..179..484H, seager_2010}, we estimate the maximum dayside brightness temperature of GJ 3929\,b to be 737 $\pm$ 14K using the effective stellar temperature from \citet{Kemmer2022} and $a/R_{\mathrm{*}}$ from \citet{xue2025jwstrockyworldsddt}. This corresponds to a brightness temperature ratio of $\mathcal{R} = 0.87\pm0.09$, which is consistent to 1$\sigma$ with both bare-rock and atmospheric models.

\section{Conclusion}\label{sec:conclusion}

We use the light-curve detrending method FN-PCA introduced in \cite{Connors2025} to fit the depths of four eclipses of GJ\,3929\,b across five observations using our open-source data reduction pipeline \verb!Erebus! with preprocessing done using \verb|Eureka!| \citep{Bell2022}. We compare the resulting eclipse depths to surface models made using \verb|JESTER| and atmosphere models made using \verb|SCARLET|. We find a dependence between eclipse depth and chosen aperture size for these data, in both our fiducial FN-PCA analysis and baseline polynomial detrending case. However, we find that the eclipse depth measurement when using our FN-PCA method is more stable under the choice of aperture size than the baseline detrending.

We find that visits 2-4 of GJ\,3929\,b follow the trend in detector-settling time presented in \cite{Connors2025}. We also find that these visits have similarly shaped detector-settling slope, which may suggest that pre-flashing the detector with the P750L filter worked to homogenize them as seen previously in \cite{fortune2025hotrockssurveyiii}. We find that visit 1 does not follow either trend, and its FN-PCA decomposition yields components that are not found in our previous analysis \citep{Connors2025}. In addition, the FN-PCA method is unable to capture the initial ramp of visit 1 without clipping the data, as the initial ramp does not match the lipped ramp shown in Figure \ref{fig:settling_pca} possibly due to nonlinear systematic effects present in this visit. The pre-flashing experiment only saw the P750L filter used for 24 minutes total prior to this discrepant observation of GJ\,3929\,b, and for under 10 minutes on target, which is the shortest amount of time out of all the other observations with seemingly homogenized detector-settling slopes. This difference may be the source of the anomalously steep settling ramp. The continuation of the pre-flashing experiment could give valuable insight into this phenomenon.

From our measured eclipse depth, $118\pm22$\,ppm, we find a dayside brightness temperature of $641^{+59}_{-64}$\,K for GJ\,3929\,b. Our measured eclipse depth rules out thick $>$1\% CO$_2$ atmospheres at $>$$3\sigma$ and agrees with all of our rocky surface models within $1\sigma$. Compared to the eclipse depth found by \cite{xue2025jwstrockyworldsddt} for the first two visits ($160^{+26}_{-27}$ ppm), the addition of two more eclipses results in an eclipse depth that is $\approx1.5\sigma$ shallower. Our analysis of only the first two visits yields a mean eclipse depth of $172\pm26$\,ppm which is well in agreement with this deeper value.  The uncertainty in the fitted eclipse depth imparted by the observational gap in visit 4 prevents us from improving the constraints on our eclipse depth as much as was anticipated. Our fitted eclipse depth is intermediate between our high- and low-albedo surface models, and we cannot rule out the possibility of a thin atmosphere on this planet. A thick CO$_2$-rich atmosphere with full heat recirculation, similar to that of Venus, is shown to be unlikely. We find up to $2.5\sigma$ variance between the eclipse depths of individual visits of GJ 3929\,b, possibly due to unmodeled instrumental or astrophysical systematic effects \citep{august_hot_2024}. This variability may also represent the true nature of the planet if it is not truly tidally locked \citep{2024A&A...690A.159P} or has a transient atmosphere \citep{2023ApJ...956L..20H}. We find that this planet is not in a circular orbit, with our fiducial result showing $e\cos\omega = -0.0225\pm0.0010$.

In the broader context of the 500 hr Rocky Worlds DDT, the results of this first planet highlight the possibility of getting an inconclusive answer from 15{\textmu}m photometry, as a bare-rock planet with high surface albedo is difficult to strongly distinguish from a dark bare rock or a thin CO$_2$ atmosphere. Furthermore, the observing gap during one of the eclipses highlights a potential pitfall when needing to split a long visit into separate observations. We find that additional observations of GJ\,3929\,b have lessened the confidence with which we can rule out certain atmospheric scenarios. We look forward to future analyses of this data using different pipelines to gauge the robustness of our conclusions and to potential future collaborations comparing data analysis methods for these tricky data sets, as was done with the Spitzer Space Telescope \citep{2016AJ....152...44I}.

\section*{Acknowledgments}

We thank the anonymous referee for their thorough review and helpful suggestions that improved this work. This work is based on observations made with the NASA/ESA/CSA James Webb Space Telescope. The data were obtained from the Mikulski Archive for Space Telescopes at the Space Telescope Science Institute, which is operated by the Association of Universities for Research in Astronomy, Inc., under NASA contract NAS 5-03127 for JWST. These observations are associated with program DD 9235. The specific JWST/MIRI observations are available at Eclipse 1 DOI: \href{https://dx.doi.org/10.17909/c2jx-a977}{doi:10.17909/c2jx-a977}, Eclipse 2 DOI: \href{https://dx.doi.org/10.17909/39mk-t407}{doi:10.17909/39mk-t407}, Eclipse 3 DOI: \href{https://dx.doi.org/10.17909/n0c3-8685}{doi:10.17909/n0c3-8685}, and Eclipse 4 DOI: \href{https://dx.doi.org/10.17909/5476-sk12}{doi:10.17909/5476-sk12}. 

We thank the Rocky Worlds DDT Core Implementation Team Leads (N. Espinoza and H. Diamond-Lowe) and their team for providing these data to the community with no exclusive access period. We thank \cite{xue2025jwstrockyworldsddt} for publishing RV results that enabled our analysis of the full JWST dataset. This research has made use of the NASA Exoplanet Archive, which is operated by the California Institute of Technology, under contract with the National Aeronautics and Space Administration under the Exoplanet Exploration Program. N.C., C.M., and P.-A.R. acknowledge financial support from the University of Montreal. L.D., C.M., and P.-A.R. acknowledge support from the Natural Sciences and Engineering Research Council (NSERC) of Canada. C.M. further acknowledges financial support from Jean-Marc Lauzon. L.D. also acknowledges support from the Trottier Family Foundation and the University of Waterloo. N.C., C.M., L.D., P.-A.R., and B.B. acknowledge support from the Canadian Space Agency. This work was made with the support of the Institut Trottier de Recherche sur les Exoplanetes (IREx).

We used the following code resources in our data analysis: \verb!astropy! \citep{astropy:2013, astropy:2018, astropy:2022}, \verb!astroquery! \citep{astroquery}, \verb!emcee! \citep{Foreman_Mackey_2013}, \verb!NumPy! \citep{harris2020array_numpy}, \verb!Eureka!! \citep{Bell2022}, \verb!batman! \citep{batman}, \verb!sklearn! \citep{scikit-learn}, \verb!matplotlib! \citep{matplotlib}, \verb!jwst! \citep{jwst_pipeline}, \verb!h5py! \citep{collette_python_hdf5_2014}, \verb!SciPy! \citep{2020SciPy-NMeth}, \verb!corner! \citep{corner}, \verb!photutils! \citep{larry_bradley_2024_13989456}, \verb!spelunker! \citep{Deal2024}, \verb!uncertainties!, and \verb!pydantic!.

\FloatBarrier

\newpage

\appendix

\begin{figure}[h]
    \centering
    \includegraphics[width=0.49\linewidth]{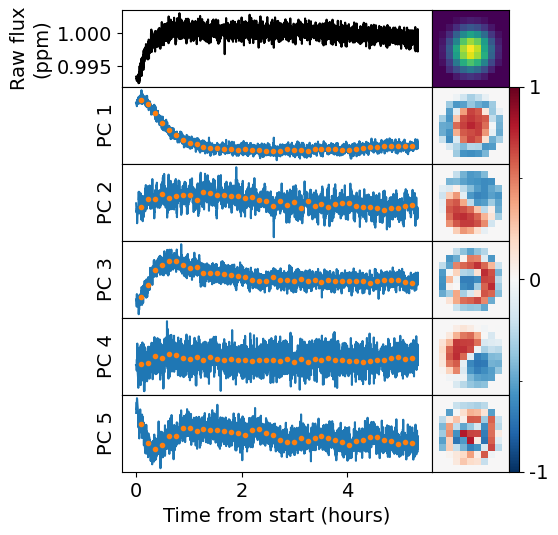}
    \includegraphics[width=0.49\linewidth]{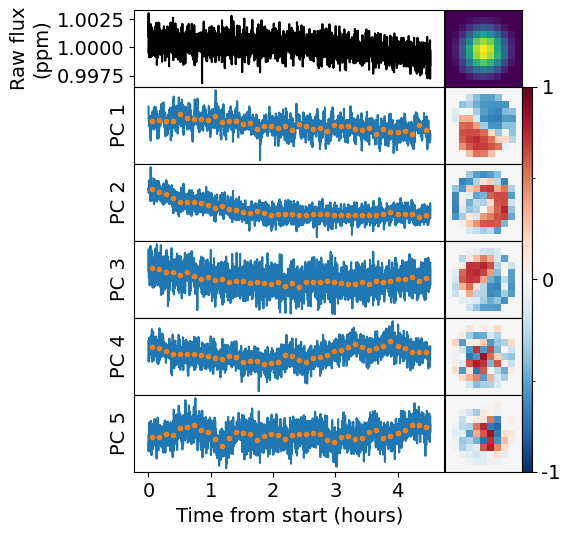}
    \caption{The frame-normalized principal component eigenvalues (left column) and eigenimages (right column) compared to the raw light curve and point-spread function (top row) for visit 1 when not trimming any data (left plot) and when trimming the first 500 integrations (right plot). The components found after trimming the first 500 integrations are used in our lightcurve fitting. We include 40 binned data points to aid in visualization of the eigenvalue time-series.}
    \label{fig:appendix_pca_visit1}
\end{figure}

\begin{figure}[h]
    \centering
    \includegraphics[width=0.49\linewidth]{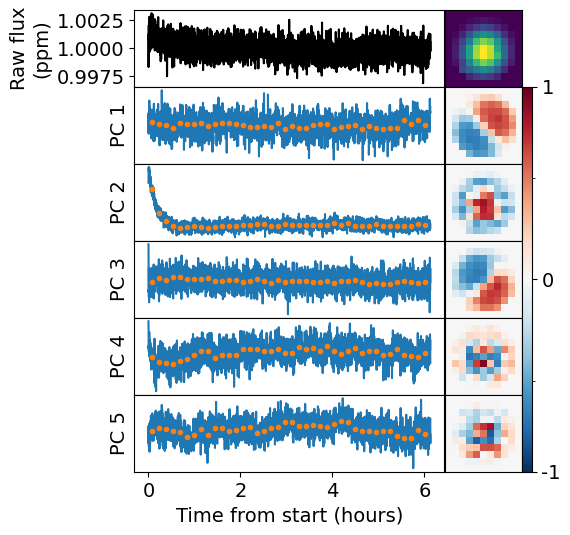}
    \includegraphics[width=0.49\linewidth]{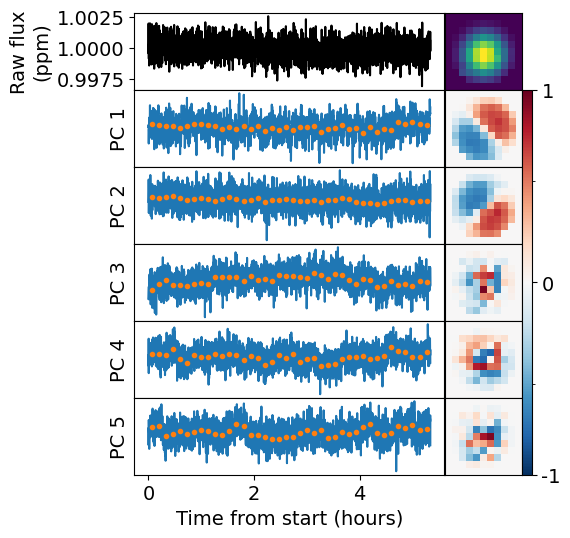}
    \caption{The frame-normalized principal component eigenvalues (left column) and eigenimages (right column) compared to the raw light curve and point-spread function (top row) for visit 2 when not trimming any data (left plot) and when trimming the first 500 integrations (right plot). The components found after trimming the first 500 integrations are used in our lightcurve fitting.}
    \label{fig:appendix_pca_visit2}
\end{figure}

\begin{figure}[h]
    \centering
    \includegraphics[width=0.49\linewidth]{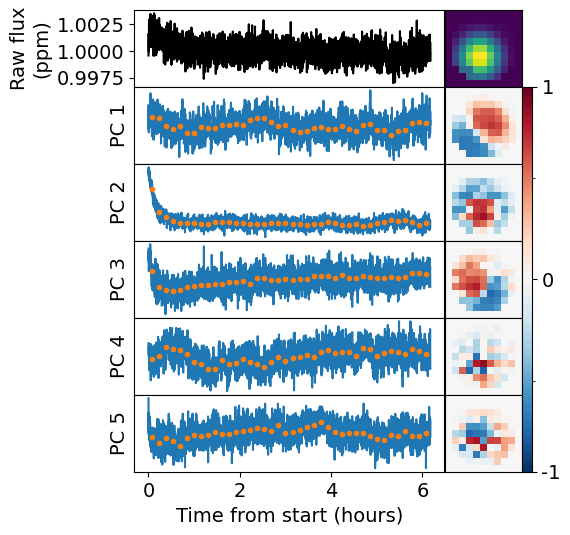}
    \includegraphics[width=0.49\linewidth]{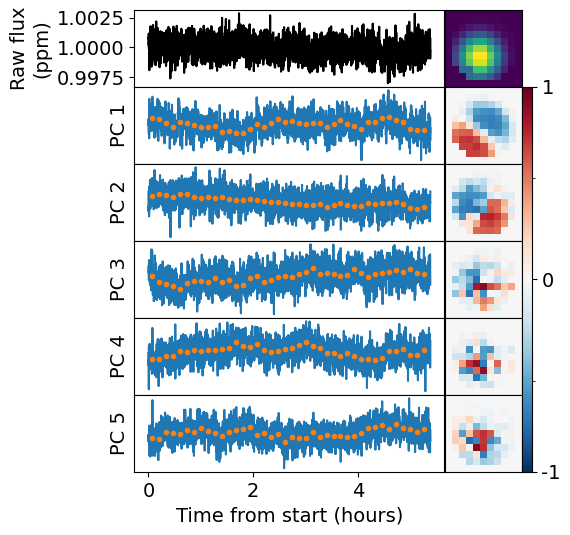}
    \caption{The frame-normalized principal component eigenvalues (left column) and eigenimages (right column) compared to the raw light curve and point-spread function (top row) for visit 3 when not trimming any data (left plot) and when trimming the first 500 integrations (right plot). The components found after trimming the first 500 integrations are used in our lightcurve fitting.}
    \label{fig:appendix_pca_visit3}
\end{figure}

\begin{figure}[h]
    \centering
    \includegraphics[width=0.49\linewidth]{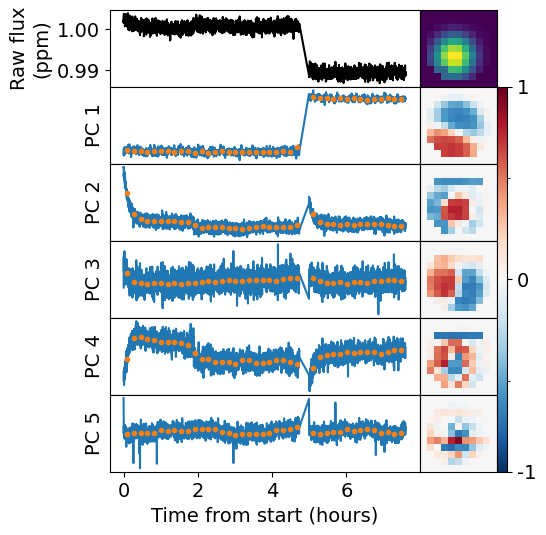}
    \\
    \includegraphics[width=0.49\linewidth]{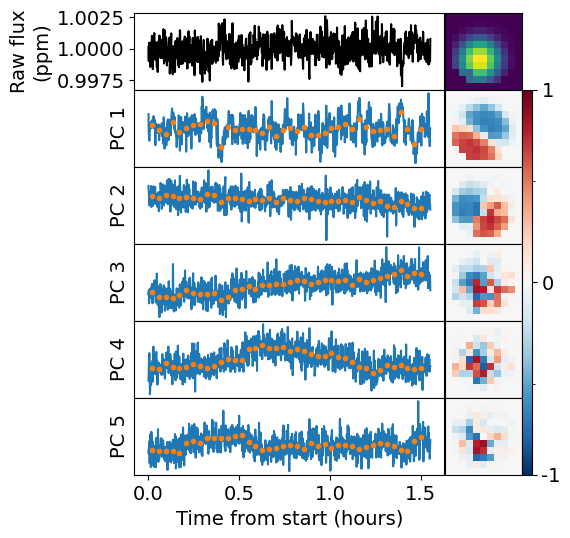}
    \includegraphics[width=0.49\linewidth]{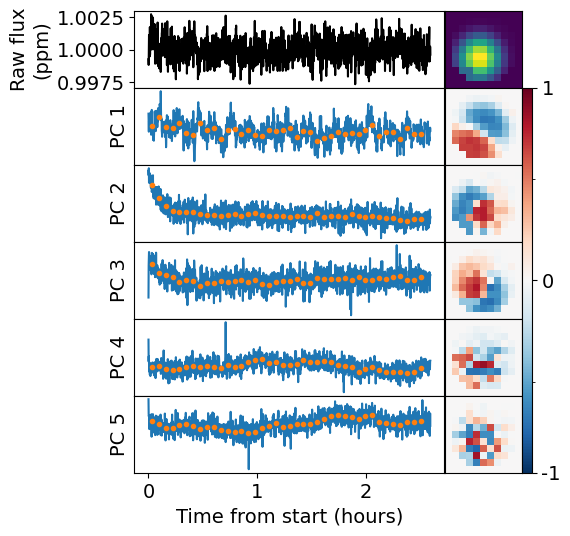}
    \caption{The frame-normalized principal component eigenvalues (left column) and eigenimages (right column) compared to the raw light curve and point-spread function (top row) for visit 4. The upper plot is the entire visit split across two observations when not trimming any data. The bottom plots are split across two observations and show the components used when fitting the lightcurve. The bottom left plot has the first 1500 integrations trimmed to remove the detector settling slope and a suspected tilt event. No data is trimmed from the bottom right plot as the beginning of the observation is near the expected time of eclipse.}
    \label{fig:appendix_pca_visit4}
\end{figure}

\begin{figure}
    \centering
    \includegraphics[width=0.45\linewidth]{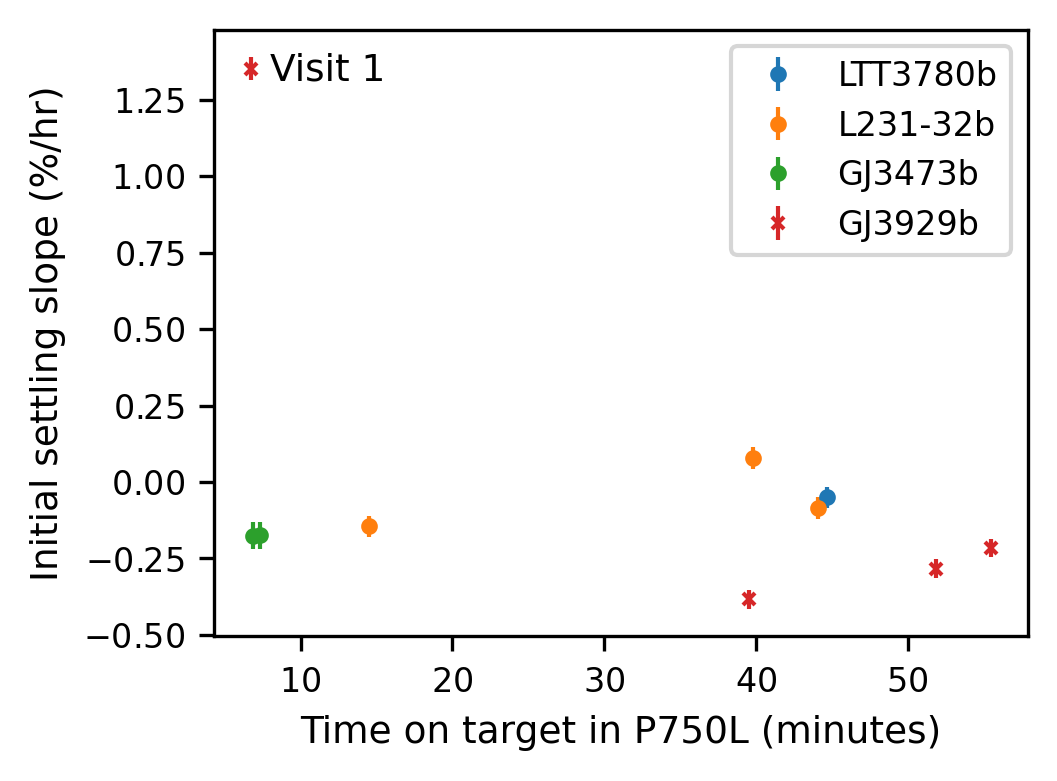}
    \includegraphics[width=0.45\linewidth]{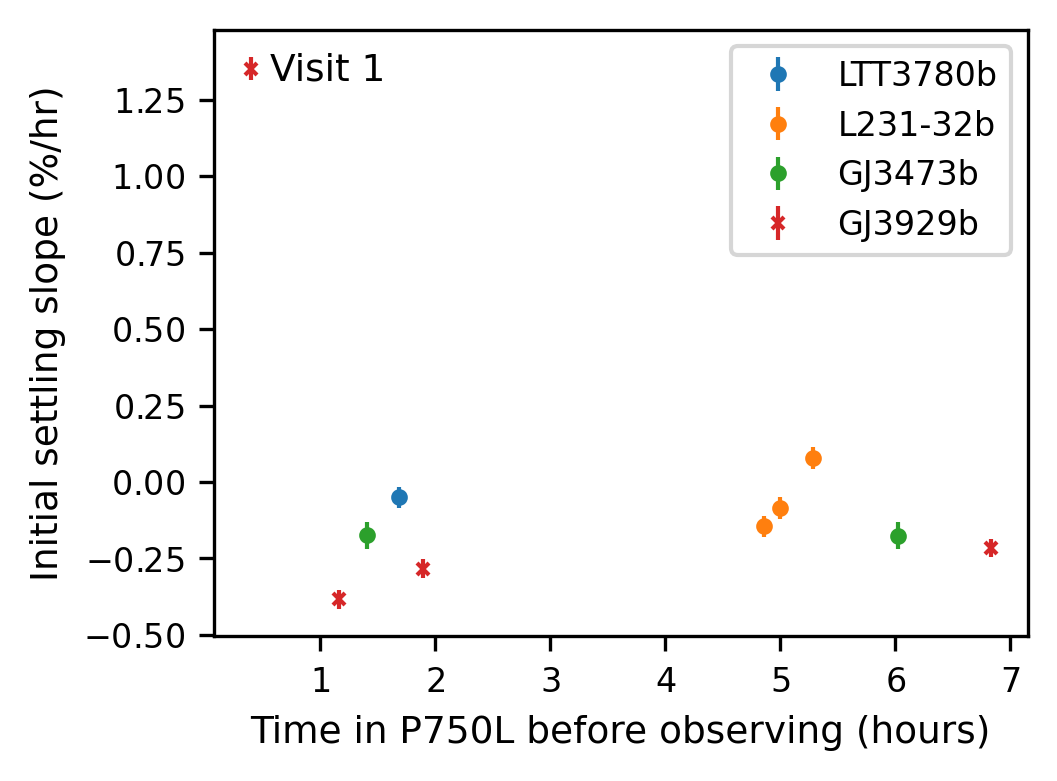}
    \caption{We compare the size of the detector settling slope (defined by fitting a linear slope to the first 30 minutes of data) to the length of time spent with the P750L filter prior to switching to F1500W, both on target (left) and in total (right), for the four eclipses of GJ\,3929\,b and the relevant targets of the Hot Rocks Survey described in \cite{fortune2025hotrockssurveyiii}. We find that all but one visit of GJ\,3929\,b have detector settling slopes similar to the other previously observed targets. The discrepant visit of GJ\,3929\,b spends a similar amount of time on target with P750L as two visits of a previous target, GJ\,3473\,b, yet does not have a homogenized slope. If we consider the total length of time with P750L before the telescope has finished slewing to the target, we see that the first visit of GJ\,3929\,b spends the lowest time overall using this filter as part of the pre-slewing experiment, while all other observations spend over 1 hour in this filter prior to starting.}
    \label{fig:appendix_p750l}
\end{figure}

Here we provide additional plots related to the detrending of these data and the systematic effects therein. For each visit of GJ\,3929\,b we plot the FN-PCA eigenvalues and eigenimages with and without trimming any data (Figures \ref{fig:appendix_pca_visit1}, \ref{fig:appendix_pca_visit2}, \ref{fig:appendix_pca_visit3}, \ref{fig:appendix_pca_visit4}). For GJ\,3929\,b and the relevant observations from the Hot Rocks Survey \citep{fortune2025hotrockssurveyiii} we plot a comparison of detector-settling slope sizes based on the amount of time spent in the P750L filter prior to the start of the observation (Figure \ref{fig:appendix_p750l}).

\FloatBarrier

\bibliography{references}



\end{document}